\documentclass[
  notitlepage,
  twocolumn,
  preprintnumbers,
  longbibliography,
  superscriptaddress,
  aps,
  prb,
  amsmath,
  amssymb,
]{revtex4-2}

\usepackage{graphicx,subfigure,epsfig}
\usepackage[normalem]{ulem}
\usepackage{dcolumn}
\usepackage{amssymb}
\usepackage{times}
\usepackage{amsmath}
\usepackage{amsfonts}
\usepackage{mathrsfs}
\usepackage{setspace}
\usepackage{latexsym}
\usepackage{bbm}
\usepackage{float}
\usepackage{placeins}
\usepackage{flafter}
\usepackage{bm}
\usepackage{epstopdf}
\usepackage{color}
\usepackage{multirow}
\usepackage{physics} 
\usepackage[export]{adjustbox}

\usepackage{braket}
\usepackage{mathtools}
\usepackage{dsfont}

\usepackage{nicematrix} 
\usepackage{tikz} 
\usetikzlibrary{decorations.pathreplacing}

\usepackage{soul} 

\def \be {\begin{eqnarray}}
\def \ee {\end{eqnarray}}

\newtheorem{theorem}{Theorem}[section]
\newtheorem{definition}{Definition}[section]

\newcommand{\bk}{{\bf k}}

\newcommand{\br}{{\bf r}}
\newcommand{\bR}{{\bf R}}
\newcommand{\bs}{{\bf s}}

\newcommand{\bmm}{{\bf m}}

\usepackage{dsfont}

\usepackage{xcolor}

 \usepackage{simpler-wick} 

\newcommand{\bea}{\begin{equation} \begin{aligned}}
\newcommand{\eea}{\end{aligned} \end{equation} }

\newcommand{\bpm}{\begin{pmatrix}}
\newcommand{\epm}{\end{pmatrix}}

\renewcommand{\Tr}{\text{Tr}}

\usepackage{xr}
\usepackage[pdftex,breaklinks,bookmarks=false,colorlinks,linkcolor=blue,citecolor=blue,urlcolor=blue,hypertexnames=false]{hyperref}

\usepackage{enumitem} 

\usepackage{orcidlink}

\begin{document}

\title{Expressivity of two-dimensional Gaussian fermion tensor network states}
\author{Pok Man Tam \orcidlink{0000-0001-9620-4689}}
\affiliation{Leinweber Institute for Theoretical Physics, University of Chicago, Chicago, IL 60637, USA}
\affiliation{Princeton Center for Theoretical Science, Princeton University, Princeton, NJ 08544, USA}%
\author{William Musk \orcidlink{0009-0008-5569-6064}}
\affiliation{%
Department of Physics and Astronomy, University of Pittsburgh, Pittsburgh, PA 15213, USA
}%
\author{Zhehao Dai \orcidlink{0000-0002-3789-5683}}
\affiliation{%
Department of Physics and Astronomy, University of Pittsburgh, Pittsburgh, PA 15213, USA
}%

\date{\today}

\begin{abstract}
Tensor network states (TNS) provide a family of efficiently compressed approximations to physically relevant many-body quantum states. An important quantity that characterizes the computational resources required by a TNS is its bond dimension. While in one dimension an area law for the R{\'e}nyi-0 entropy guarantees an exact matrix product state representation with a bond dimension independent of the system size, an analogous statement does not hold for general states in two dimensions (2D). This leaves open the question of how the expressivity of 2D TNS is related to the entanglement properties of the states they represent. In this work, we focus on two families of 2D translation-invariant Gaussian fermionic states: single-band BCS superconductors and half-filled two-band insulators (including semimetals). For each family, we prove that an area law for the R{\'e}nyi-0 entropy is equivalent to having an exact Gaussian fermionic TNS representation with a finite bond dimension determined by the R{\'e}nyi-0 entropy.
We also establish equivalent characterizations in terms of compactly supported Wannier-type orbitals and the momentum-space wavefunction, relating the spatial extent of these orbitals and the structure of the wavefunction to the required bond dimension.
\end{abstract}

\maketitle

\section{Introduction}\label{sec:intro}
Tensor network states (TNS) 
offer an efficiently compressed approximation for a large class of physically relevant many-body states~\cite{cirac2021matrix,Xiang2023}.
In one dimension (1D), matrix product states (MPS) provide the variational framework for the density-matrix renormalization group (DMRG) method whose success has made accurate ground-state and dynamical calculations possible for a wide range of one-dimensional systems~\cite{white1992density,schollwock2011density}.
For two-dimensional (2D) systems, projected entangled-pair states (PEPS) extend this framework by connecting tensors along the bonds of the 2D lattice~\cite{nishino2001twodimensional,verstraete2004renormalization,PhysRevLett.96.220601,murg2007variational,full_update_1}.
Advances in PEPS methods have established them as a powerful tool for investigating 2D quantum matter with successful applications to frustrated magnets~\cite{Murg2009Frustrated,CorbozMila2013,Wang2013J1J2,CorbozMila2014,Liu2022J1J2}, quantum spin liquids and topological phases~\cite{Poilblanc2012RVB,Poilblanc2015Chiral,Chen2020CSL,Chen2021Topological,chiral_iPEPS_I,YangCorboz2026}, and competing stripe and superconducting orders in strongly correlated fermion models~\cite{Corboz2011Stripes,Corboz2014,Zheng2017,Ponsioen2019Stripes,Zhang2025Hubbard,Liu2025Hubbard}.
Their applications also extend to excitation spectra~\cite{planewave1,planewave2,planewave3,planewave4}, real-time dynamics~\cite{Czarnik2019TimeEvolution,Kshetrimayum2020,fermion_dynamics_I,lin2021efficient,boson_dynamics_I,Dai2025IsoTNS}, and benchmarks for quantum simulation experiments~\cite{boson_dynamics_I,ORourke2023Rydberg,Kim2023Utility,Tindall2024Eagle,Begusic2024Converged,Patra2024,Zhang2025Google}.
Advances in contraction, optimization, and Monte Carlo sampling continue to improve their computational reach~\cite{corner_transfer_matrix,Corboz2016Optimization,Vanderstraeten2016Gradient,Liu2017Gradient,Liao2019Differentiable,SchindlerJermyn2020Contraction,Zhang2026QR,Zhang2026Preconditioning,WuDai2026VMC,du2026vectorized}.

An important step towards large-scale TNS simulations is to understand their \textit{expressivity}: which classes of many-body states, characterized by their physical properties, can be accurately approximated by a given TNS structure at a prescribed bond dimension.
The bond dimension is the size of the auxiliary indices connecting tensors; increasing it enlarges the variational family but also raises the computational cost.
In 1D, entanglement provides a particularly useful way to characterize MPS expressivity.
An MPS of bond dimension $D$ supports an entanglement entropy of at most $\log D$ across a single-bond cut ~\cite{schollwock2011density}.
Conversely, area-law bounds on R{\'e}nyi entropies give sufficient conditions for efficient MPS approximation~\cite{VerstraeteCirac2006,Schuch2008Entropy}.
In 2D, gapped topological states such as toric-code and other quantum-double ground states~\cite{Schuch2010Topology} and string-net ground states~\cite{Gu2009StringNet,Buerschaper2009StringNet,stringnet_isotns} admit exact PEPS representations with bond dimensions independent of system size.
Such representations are also known for the short-range kagome resonating-valence-bond spin liquid~\cite{Schuch2012RVB}; all these states obey an entanglement area law by construction~\cite{cirac2021matrix,Xiang2023}. However, area-law entanglement alone does not guarantee an efficient global PEPS approximation: Ge and Eisert constructed families of states satisfying area laws for all R{\'e}nyi entropies that cannot be globally approximated by PEPS with a polynomial-size classical description~\cite{GeEisert2016}. This motivates a closer examination of how entanglement and other physical properties constrain PEPS expressivity and the bond dimensions required to represent physically relevant 2D states~\cite{Cirac2019OpenProblems,cirac2021matrix}.

In this work, we focus on 2D translation-invariant Gaussian fermionic tensor network states (GfTNS)~\cite{PhysRevA.81.052338,GfTNS_chiral,gu2010grassmann,Dubail_Read,Hackenbroich2020,He2024StackedTree} for which analytical progress can be made in addressing the above expressivity question. GfTNS form a subclass of fermionic TNS in which each local tensor can be viewed as a Gaussian state of the physical fermions and the virtual fermions~\cite{PhysRevA.81.052338}. With $n$ complex virtual fermion modes per bond, the corresponding bond dimension is $2^n$. The study of fermionic TNS is motivated by the central role of itinerant electrons in quantum materials whose phases include Fermi-liquid metals, Dirac semimetals, superconductors, and insulators. Gaussian states provide free-fermion and mean-field starting points for studying these phases as well as useful benchmarks for developing tensor-network methods for interacting electronic systems~\cite{PhysRevA.81.052338,TN_Fermi_Surface,Li2023GaussianPEPS,Yang2023ProjectedDWave,Ono2026FrustrationFree,Masaoka2026FrustrationFree,Sengoku2025Quasilocal}. At the same time, the relative simplicity of Gaussian states allows a detailed analytical investigation of the relation between entanglement and tensor-network expressivity~\cite{Dubail_Read,Finite_Ent_Scaling_2023}. The goal of this work is to analytically establish a precise, quantitative connection between the R{\'e}nyi-0 entropy and the minimal bond dimension required for a GfTNS representation. 
Given the counterexamples for general TNS found by Ge and Eisert \cite{GeEisert2016}, this connection is nontrivial and reassuring.
Below, we first provide a concise overview for our key results. 

\subsection{Overview}
We focus on two classes of systems: one consists of single-band BCS superconductors and another consists of $U(1)$ charge conserved two-band insulators at half-filling (i.e., with one filled band, and hence also including semimetals). We will provide a concise review on the structure of 2D GfTNS in Sec.~\ref{sec: GfTNS review}. Here let us first summarize our main results regarding the TNS representation of a generic single-band BCS state. A single-band translation-invariant BCS state is generically represented as
\begin{equation}\label{eq: BCS Gaussian form}
\begin{split}
    \ket{\psi_p} &\propto \prod_\bk e^{\frac{1}{2}M^p(\bk)b^\dagger_{-\bk} b^\dagger_{\bk}}\ket{0}_p \\
    &= \sideset{}{'}\prod_{\bk}(1+M^p(\bk)b^\dagger_{-\bk} b^\dagger_{\bk}) \ket{0}_p
\end{split}
\end{equation}
where $\ket{0}_p$ is the physical vacuum state, $b^\dagger_\bk$ creates a fermion at momentum $\bk$, and $M^p(\bk)=-M^p(-\bk)$ (by virtue of the fermionic anti-symmetry) parametrizes the BCS state and is referred to as the momentum-space wavefunction. In the last equation, $\prod_{k}^{'}$ indicates that we choose one of the momenta of every time-reversal pair $\bk$, $-\bk$. To characterize the expressivity of GfTNS, we will prove in Sec.~\ref{sec: single-band BCS}  that the following four properties are equivalent for Gaussian states in Eq.~\eqref{eq: BCS Gaussian form}:

\begin{enumerate}[label=(\textbf{P\arabic*}):, ref=(P\arabic*)]
    \item \textit{Area law of R{\'e}nyi-0 entropy:} For any region $A$ of size $l_x\times l_y$ sites, the R{\'e}nyi-0 entanglement entropy satisfies an upper bound, $S_0(A)\le 2n_xl_y + 2n_yl_x$. \label{itm: P1}
    \item \textit{Compactly supported Wannier-type orbital:} There is a compactly supported Wannier-type operator of size $(2n_x+1)\times (2n_y+1)$. \label{itm: P2}
    \item \textit{Momentum-space wavefunction:} $M^p(\bk)$ can be represented as a ratio of trigonometric polynomials  $u(\bk)=u(-\bk)$ and $v(\bk)=-v(-\bk)$ of degree $(n_x,n_y)$. \label{itm: P3}
    \item \textit{Tensor network representation:} The state is represented by a Gaussian fermion TNS (or its particle-hole conjugate) with $n_x$ and $n_y$ virtual complex fermions on each horizontal and vertical leg of the tensor, respectively; correspondingly, the horizontal (vertical) bond dimension is $2^{n_x}$ ($2^{n_y}$). \label{itm: P4}
\end{enumerate}
We provide definitions for terminologies such as R{\'e}nyi-0 entropy, Wannier-type orbitals and the degree of trigonometric polynomials in Sec.~\ref{sec: GfTNS review}. Later on in Sec.~\ref{sec: U(1) TNS review}, we describe the structure of $U(1)$-symmetric GfTNS, and in Sec.~\ref{sec: U(1) insulator, expressivity} we correspondingly establish  four equivalent statements related to the GfTNS expressivity for $U(1)$ symmetric two-band Gaussian states which include generic two-band insulators.

Overall, our work establishes, for two classes of fermionic many-body states, that an area-law entanglement does provide a sufficient condition on the minimal bond dimension for their GfTNS representation.
In particular, the area-law bound determines the size of the maximally localized Wannier-type orbital, which in turn constrains the explicit form of the many-body wavefunction, and subsequently determines the minimal bond dimension for its TNS representation. 

\section{Gaussian fermion Tensor network}\label{sec: GfTNS review}
In this section, we review the background for constructing Gaussian fermionic tensor network states and discuss the related notion of trigonometric polynomials and Wannier-type orbitals as well as entanglement properties of Gaussian states. The following discussion serves to motivate the connection between \ref{itm: P1}-\ref{itm: P4}. We rigorously prove the equivalence between these four properties later in Sec.~\ref{sec: single-band BCS}.

\subsection{GfTNS construction}\label{subsec: GfTNS formalism}

\begin{figure}[t]
    \centering
    \resizebox{\columnwidth}{!}{\includegraphics[]{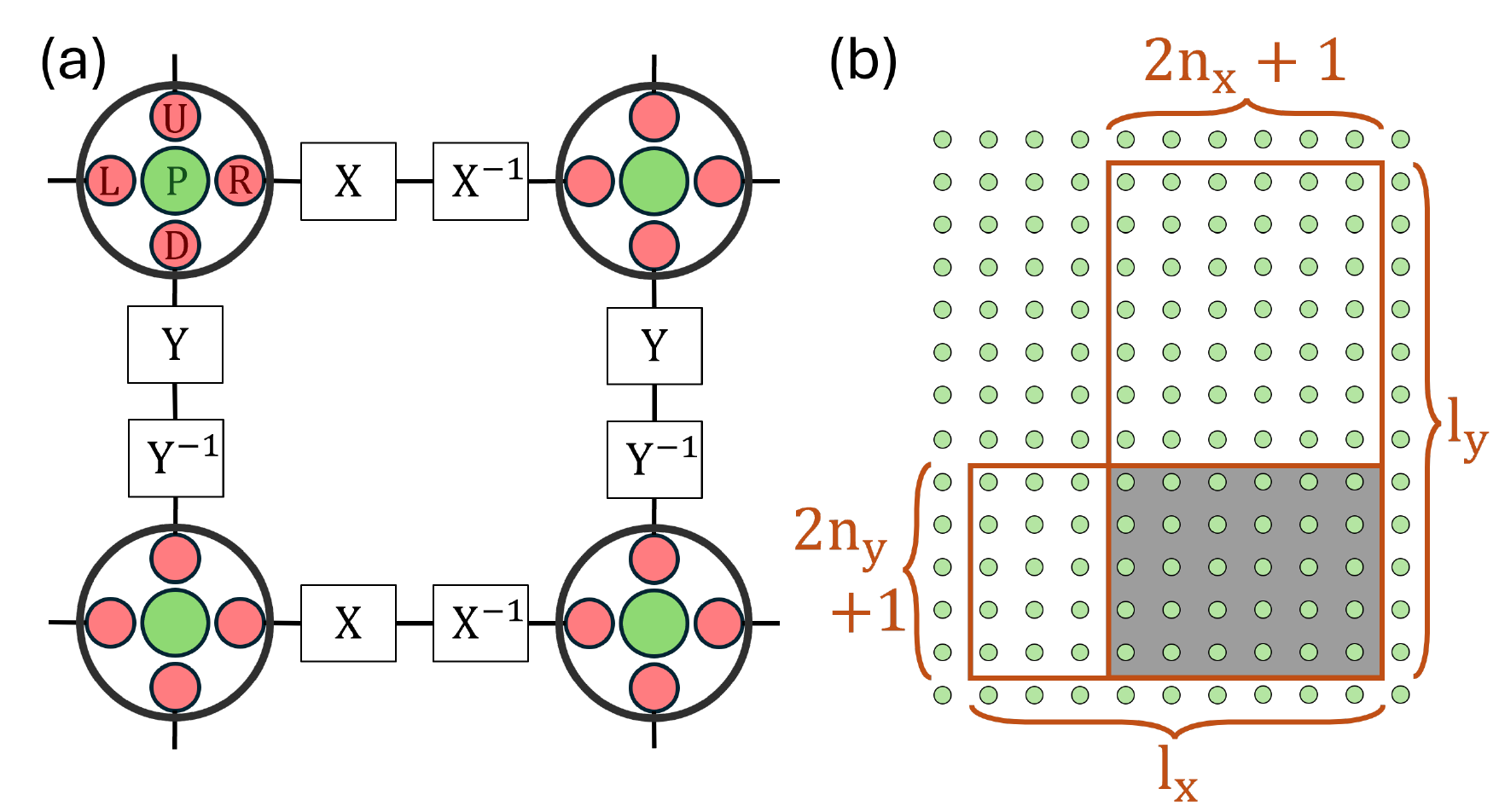}}
    \caption{(a) Structure of a 2D tensor network state. At each lattice site, there is a local tensor (large black circle) with a physical leg labeled $p$ and four virtual legs, labeled $L,R,D,U$, respectively. For a GfTNS, physical and virtual complex fermionic modes at each site are coupled to form a Gaussian fiducial state. The GfTNS is constructed by projecting neighboring virtual states onto maximally entangled pair states, as depicted by the solid straight lines. There are gauge redundancies from the fact that consistent rotations of the virtual basis leave the tensor contraction, and hence the physical state, unchanged.  (b) Pictorial aid for proving that an area law bound for the R{\'e}nyi-0 entropy implies the existence of a compactly supported Wannier-type orbital. The size of the maximally localized Wannier-type orbital is related to the R{\'e}nyi-0 entropy.}
    \label{fig: GfTNS structure}
\end{figure}

Here we review the formalism for constructing a Gaussian fermionic tensor network state (GfTNS)~\cite{PhysRevA.81.052338,GfTNS_chiral, gu2010grassmann, Dubail_Read, Hackenbroich2020, cirac2021matrix, Finite_Ent_Scaling_2023} and provide the necessary background for understanding the main results of our work regarding its expressivity. We begin by constructing the local Hilbert space on a 2D square lattice (the 1D case can be straightforwardly deduced) where each lattice site, labeled by its location $\br$, contains both physical and virtual complex fermionic modes. As we will see, a physical TNS state $\ket{\psi_p}$ is obtained by projecting a product state $\ket{\psi_{vp}}$ in the full Hilbert space onto a certain virtual state $\ket{\psi_v}$ that mediates entanglement. We denote the creation operator for a physical mode at site $\br$ by $b^\dagger_{(\br, \beta)}$, where $\beta$ $=1,...,n_p$ is the physical orbital index per site (with $n_p\le 2$ in most of this work). The creation operator for a virtual mode is denoted by $v^\dagger_{(\br, \nu)}$, where $\nu$ labels the internal virtual orbitals. In 2D, virtual modes on each site are organized into four sectors (sometimes referred to as ``legs"): $L$ (left), $R$ (right), $D$ (down) and $U$ (up), see Fig.\ref{fig: GfTNS structure}(a). The number of virtual modes in the $L$ or $R$ sector is denoted as $n_x$, while that in the $D$ or $U$ sector is denoted as $n_y$. Our work concerns how $n_{x,y}$ depends on the physical and entanglement properties of Gaussian states.

We use a pure-state formalism of the GfTNS equivalent to the Grassmann-integral formalism discussed in Ref.~\cite{gu2010grassmann,Finite_Ent_Scaling_2023} but stated in terms of virtual state projection to make connection to general fermionic TNS.
The GfTNS construction begins with a Gaussian state involving both physical and virtual fermionic modes (uniformly denoted by the creation operator $a^\dagger$):
\begin{equation}\label{eq: fiducial state without U(1)}
    \ket{\psi_{vp}} \propto e^{\frac{1}{2}\mathscr{N}_{ij} a^\dagger_ia^\dagger_j} \ket{0}_v\ket{0}_p
\end{equation}
where indices $i, j$ have a one-to-one correspondence to either $(\br, \beta)$ for a physical mode or $(\br, \nu)$ for a virtual mode, and the Einstein summation convention is adopted throughout this work. The physical and virtual vacuum states are denoted as $\ket{0}_p$ and $\ket{0}_v$, respectively. Importantly, the $\mathscr{N}$-matrix is assumed to be \textit{diagonal} in the position      basis,
\begin{equation}
    \mathscr{N}_{(\br, \alpha),(\br',\alpha')} =N_{\alpha,\alpha'}\delta_{\br,\br'},
\end{equation}
where $N_{\alpha,\alpha'}$ specifies the local fiducial state \cite{GfTNS_chiral, Hackenbroich2020} in the space of physical and virtual fermions at each site. This local fiducial state defines a local tensor where the tensor elements are wavefunction amplitudes of this fiducial state in the Fock basis. 
This is schematically represented by the large black circle in Fig.~\ref{fig: GfTNS structure}(a). 
It is convenient to arrange the site-independent $N$-matrix into the following form:
\begin{equation}
    N=\begin{pmatrix}
    N^{pp} & N^{pv} \\ N^{vp} & N^{vv}
\end{pmatrix}
\end{equation}
where $N^{pp}$ specifies the local couplings (or correlation) among physical modes, $N^{vv}$ specify the local couplings among virtual modes, while $N^{pv}$ and $N^{vp}$ specify the local couplings between physical and virtual modes. By virtue of the fermionic anti-commutation, we have $N^{pp}=-(N^{pp})^T$, $N^{vv}=-(N^{vv})^T$ and $N^{vp}=-(N^{pv})^T$. 

An important next step of the construction is the virtual-state projection. Introducing a reference virtual state as 
\begin{equation}\label{eq: reference virtual state}
    \ket{\psi_v} \propto e^{\frac{1}{2}M_{ij}v^\dagger_i v^\dagger_j} \ket{0}_v,
\end{equation}
the physical TNS $\ket{\psi_p}$ is then obtained by projecting the virtual component onto the reference virtual state:
\begin{equation}\label{eq: physical state projection}
    \ket{\psi_p} :=\braket{\psi_v|\psi_{vp}} \propto e^{\frac{1}{2}M^p_{ij}b^\dagger_i b^\dagger_j} \ket{0}_p.  
\end{equation}
It is straightforward to work out, as detailed in Appendix~\ref{appendix: virtual state}, the relation between the physical state wavefunction (as specified by $M^p$) and all the TNS parameters:
\begin{equation}\label{eq: real_space_relation} 
    M^p = \mathscr{N}^{pp} + \mathscr{N}^{pv}[\mathscr{N}^{vv}+(M^\dagger)^{-1}]^{-1} (\mathscr{N}^{pv})^{T}.
\end{equation}
Following the idea of projected entangled pair states \cite{verstraete2004renormalization, PhysRevLett.96.220601, PhysRevA.81.052338, cirac2021matrix, Finite_Ent_Scaling_2023}, the virtual state $\ket{\psi_v}$ being projected onto is composed of maximally entangled pair states for all nearest-neighbor virtual modes. More concretely, the virtual-state wavefunction is specified by an antisymmetric $M$ such that in a horizontal (vertical) bond the $R$ ($U$) virtual modes at $\br$ are diagonally coupled to the $L$ ($D$) virtual modes at $\br+\hat{x}$ ($\br+\hat{y}$):
\begin{equation}\label{eq: real-space M}
    M_{(\br, \nu_R), (\br+\hat{x},\nu_L)} = \mathds{1}_{n_x}, \quad M_{(\br, \nu_U), (\br+\hat{y},\nu_D)}=\mathds{1}_{n_y},
\end{equation}
while any remaining unspecified matrix elements for $M$ vanish. Correspondingly, the horizontal and vertical bond dimensions (i.e., the dimension of the virtual Hilbert space on each bond following the projection) are $2^{n_x}$ and $2^{n_y}$, respectively. This virtual-state correlation is represented by the black solid straight lines in Fig.~\ref{fig: GfTNS structure}(a). 
In passing, we note that Gaussian fermionic TNS form a subclass of general fermionic TNS in which the local fiducial state involving physical and virtual fermions are restricted to be Gaussian.
To describe interacting fermionic systems, one can adopt non-Gaussian fiducial states while retaining the same virtual-state projection construction~\cite{PhysRevA.81.052338}.

For our purpose of investigating translation-invariant states, it is most convenient to work in the momentum basis, where the physical state can be represented as
\begin{equation}\label{eq: Gaussian form of BCS}
    \ket{\psi_p} \propto\prod_\bk \exp[\frac{1}{2}\sum_{\beta,\beta'=1}^{n_p}[M^p(\bk)]_{\beta\beta'}b^\dagger_{(-\bk,\beta)}b^\dagger_{(\bk,\beta')}]\ket{0}_p,
\end{equation}
with 
\begin{equation}
    [M^p(\bk)]_{\beta,\beta'} = \sum_{\br}M^p_{(\br, \beta), (0, \beta')} e^{i\bk\cdot \br}.
\end{equation}
As detailed in App.~\ref{appendix: fourier}, the real-space relation in Eq.~\eqref{eq: real_space_relation} corresponds to the following relation in the momentum space: 
\begin{equation}\label{eq: k-space formula for M^p}
    M^p(\bk) = N^{pp}+N^{pv}[N^{vv}+M(\bk)]^{-1}(N^{pv})^T,
\end{equation}
where $N^{pp}$ has dimension $n_p \times n_p$ (with $n_p$ the number of physical orbitals per site), $N^{pv}$ has dimension $n_p \times (2n_x+2n_y)$, and $N^{vv}$ has dimension $(2n_x+2n_y) \times (2n_x+2n_y)$. Using the real-space representation in Eq.~\eqref{eq: real-space M}, we obtain
\begin{equation}\label{eq: Mk_2D}
    M(\bk) = \begin{pNiceArray}[first-row,first-col]{cccc}
    & L & R & D & U\\
  L & & -e^{ik_x} \mathds{1}_{n_x} & & \\
  R & e^{-ik_x} \mathds{1}_{n_x} & & & \\
  D & & & & -e^{ik_y} \mathds{1}_{n_y}\\
  U & & & e^{-ik_y} \mathds{1}_{n_y}&
    \end{pNiceArray},
\end{equation}
where for clarity we have also indicated on the sides the virtual-mode basis sectors under which $M(\bk)$ is expressed. For consistency, we use the same ordering for the virtual-mode basis $(\nu_L, \nu_R, \nu_D, \nu_U)^T$ throughout the entire work for matrices involved in the TNS construction. For 1D states, with $n$ virtual modes in each of the $L$ or $R$ sector,
\begin{equation}\label{eq: Mk_1D}
    M(k) = \begin{pNiceArray}[first-row,first-col]{cc}
    & L & R \\
  L &  & -e^{ik} \mathds{1}_{n}  \\
  R & e^{-ik} \mathds{1}_{n} & 
    \end{pNiceArray},
\end{equation}
while $N^{pv}$ has dimension $n_p \times 2n$ and $N^{vv}$ has dimension $2n\times 2n$, respectively.  Notice that $M(\bk) = -[M(-\bk)]^T$, hence Eq.~\eqref{eq: k-space formula for M^p} implies $M^p(\bk) = - [M^p(-\bk)]^T$ which is consistent with the fermion anti-symmetry as seen from Eq.~\eqref{eq: Gaussian form of BCS}.  Equation~\eqref{eq: k-space formula for M^p}, relating the TNS parameters $N$ to the momentum-space wavefunction $M^p(\bk)$ of the physical state, serves as the starting point for our subsequent analysis on the expressivity of GfTNS. 

\subsubsection{Gauge redundancy}
It is important to note from Eq.~\eqref{eq: k-space formula for M^p} that the mapping from the TNS parameters $N$ to the physical state $\ket{\psi_p}$ is \textit{not} one-to-one, as different choices of $N$ can come together to form the same $M^p(\bk)$. 
This redundancy in the TNS representation is referred to as a gauge redundancy. Schematically, as the TNS can be represented in the form of Fig.~\ref{fig: GfTNS structure}(a) as tensor contraction along the virtual bonds, one can always insert an identity operator acting on the virtual Hilbert space ($XX^{-1}=\mathds{1}_{2^{n_x}}$ for the horizontal bond or $YY^{-1}=\mathds{1}_{2^{n_y}}$ for the vertical bond) without affecting the result of the tensor contraction and hence leaving the physical state unchanged. Equivalently, this identity-insertion can be interpreted as a transformation of the local tensor by conjugation, under which the virtual basis $v^\dagger_{(\br,\nu)}$ on each horizontal (vertical) leg is rotated by a Gaussian operator $e^{H_x}$ ($e^{H_y}$), which is parametrized by a quadratic $H_{\rho}$ ($\rho=x,y$):
\begin{equation}\label{eq: gauge transform, BCS case}
    H_\rho = \sum_{i,j=1}^{n_\rho}v^\dagger _i [H_\rho^{(1)}]_{ij} v^\dagger_j +v _i [H_\rho^{(2)}]_{ij} v_j+ v^\dagger _i [H_\rho^{(3)}]_{ij} v_j
\end{equation}
with $H_\rho^{(1,2)}$ being anti-symmetric and $H_\rho^{(3)}$ completely generic, $n_\rho\times n_\rho$ matrices. In 2D, gauge redundancies are thus parametrized by $2(n_x^2+n^2_y)-n_x-n_y$ number of complex parameters. 

For GfTNS without $U(1)$ symmetry, we focus on the case $n_p=1$, i.e., one physical orbital per site. 
The antisymmetric matrix $N$ has $(n_x+n_y)(2n_x+2n_y+1)$ parameters.
After eliminating gauge redundancies, the number of independent TNS parameters (denoted as $\mathcal{N}$) for such a physical state constructed with $n_x$ ($n_y$) virtual modes on each horizontal (vertical) leg is therefore
\begin{equation}\label{eq: number of indep. TNS param.}
    \mathcal{N} = 4n_xn_y+2n_x+2n_y.
\end{equation}
Similarly, for 1D GfTNS with $n$ virtual modes per leg and one physical orbital per site, we have $\mathcal{N}=2n$. Next, we return to Eq.~\eqref{eq: k-space formula for M^p} and examine the general form for $M^p(\bk)$ in more detail.

\subsection{Trigonometric polynomials}\label{subsec: TP}
Combining Eq.~\eqref{eq: k-space formula for M^p} with Eq.~\eqref{eq: Mk_2D}, it is easy to see that the matrix elements of $M^p(\bk)$ all take the form as a ratio of trigonometric polynomials \cite{Dubail_Read, GfTNS_chiral, Read2017CompactWannier}. A trigonometric polynomial (TP) is a Laurent polynomial of $e^{ik_x}$ and $e^{ik_y}$, and we refer to its largest power (taking into account the magnitude of both positive and negative powers) as the degree of the TP. In particular, for a BCS state with $n_p=1$, we can always express
\begin{equation}\label{eq: Mp as a ratio of TP}
    M^p(\bk) = \frac{v(\bk)}{u(\bk)}
\end{equation}
where $u(\bk)$ and $v(\bk)$ are parity-even and parity-odd TPs, respectively, of the following general forms:
\begin{equation}\label{eq: general TPs}
\begin{split}
    &u(\bk) = u_\mathbf{0}+\sum_{\bmm\in \mathcal{L}{(d_x,d_y)}} u_\bmm \cos(\bmm\cdot \bk), \\
    &v(\bk) = \sum_{\bmm\in \mathcal{L}{(d_x,d_y)}} v_\bmm \sin(\bmm\cdot \bk),
\end{split}
\end{equation}
where $u_\bmm, v_\bmm\in \mathbb{C}$ are complex TP coefficients. The summation above runs over the following set of non-trivial lattice points
\begin{equation}\label{eq: non-trivial lattice  points}
\begin{split}
&\mathcal{L}(d_x,d_y)
:=\\
&\left\{
\bmm=(m_x,m_y)\in\mathbb{Z}^2
\middle|
\begin{array}{l}
1\leq m_x\leq d_x, -d_y\leq m_y\leq d_y,\\
\text{or}\quad m_x=0, \quad\;\;1\leq m_y\leq d_y
\end{array}
\right\}.
\end{split}
\end{equation}
We refer to $(d_x,d_y)$ as the \textit{degree} of the TPs in Eq.~\eqref{eq: general TPs}.

It is easy to see that, for a given TP degree $(d_x,d_y)$, the parity-odd $v(\bk)$ contains $\abs{\mathcal{L}(d_x,d_y)}=2d_xd_y+d_x+d_y$ independent coefficients and the parity-even $u(\bk)$ contains $\abs{\mathcal{L}(d_x,d_y)}+1$ independent coefficients. Since $M^p(\bk)$ in Eq.~\eqref{eq: Mp as a ratio of TP} consists of a \textit{ratio} of two TPs, after eliminating one parameter for the overall scaling of both $u$ and $v$, we see that $M^p(\bk)$ is determined by $4d_xd_y+2d_x+2d_y$ complex parameters. Taking $d_x=n_x$ and $d_y=n_y$, we obtain $2\abs{\mathcal{L}}=\mathcal{N}$, an exact match between the counting of TP parameters determining $M^p(\bk)$ and the number of independent TNS parameters obtained in Eq.~\eqref{eq: number of indep. TNS param.}. This simple counting argument motivates the relation between \ref{itm: P3} and \ref{itm: P4}. While given the above discussion it is clear that \ref{itm: P4} $\implies$ \ref{itm: P3}, showing \ref{itm: P3} $\implies$ \ref{itm: P4} requires more work by examining the detailed dependence of the TP coefficients on the TNS parameters, which we discuss in Sec.~\ref{sec: P3 implies P4}.

Next, we discuss the notion of Wannier-type orbital, which is the central object in \ref{itm: P2} and serves to connect the entanglement property \ref{itm: P1} of a Gaussian state to its TNS representation \ref{itm: P4}.

\subsection{Compactly supported Wannier-type orbitals}\label{subsec: WTO, BCS}

A Wannier-type orbital (WTO) is an occupied single-particle state that is localized in real space, without requiring its lattice
translates to be orthogonal \cite{Anderson_Wannier, Marzari2012WannierReview, Dubail_Read, Read2017CompactWannier, SchindlerBernevig2021Noncompact, Masaoka2026FrustrationFree, Ono2026FrustrationFree}. For our discussion on Gaussian states without $U(1)$ symmetry, the WTO can generally be a superposition of particle and hole states. Given a Gaussian state of the form in Eq.~\eqref{eq: BCS Gaussian form}, with $M^p(\bk)=v(\bk)/u(\bk)$, it is easy to check that the following operator
\begin{equation}
    \gamma^\dagger_\bk=u(\bk)b_{-\bk}-v(\bk)b^\dagger_{\bk}
\end{equation}
creates an occupied orbital in momentum space, as
\begin{equation}\label{eq: gamma_k annihilates BCS}
    \gamma^\dagger_\bk \ket{\psi_p} \propto v(\bk)\left(b_{-\bk}b^\dagger_{-\bk}b^\dagger_\bk-b^\dagger_\bk\right)\ket{0}_p = 0, 
\end{equation}
which holds for a generic $\bk$ \footnote{While Eq.~\eqref{eq: gamma_k annihilates BCS} does not hold for inversion-symmetric $\bk$ points (e.g., $\bk=0$), as we do not require $v(\bk)$ to be parity-odd here, such $\bk$ points form only a measure-0 subset in the entire momentum-space and have no effect in our subsequent analysis upon taking the thermodynamic limit. Alternatively, we can consider anti-periodic boundary conditions to avoid these inversion-symmetric $\bk$ points.}.
The Fourier transform of $\gamma^\dagger_\bk$ thus gives an occupied orbital in real space:
\begin{equation}\label{eq: def. of WTO}
\begin{split}
    \Gamma^\dagger_\bR &= \frac{1}{\sqrt{N_s}}\sum_\bk e^{-i\bk\cdot\bR}\gamma^\dagger_\bk \\
    &=\sum_{\br}\left(\tilde{u}(\br) b_{\bR+\br} -\tilde{v}(\br)b^\dagger_{\bR+\br}\right)
\end{split}
\end{equation}
where $b^\dagger_\br = \frac{1}{\sqrt{N_s}}\sum_\bk e^{-i\bk\cdot\br } b^\dagger_\bk$, $N_s$ is the number of physical sites, and 
\begin{equation}\label{eq: def. of u and v in real space}
    \tilde{u}(\br) = \frac{1}{N_s}\sum_\bk e^{i\bk\cdot\br} u(\bk),\;\tilde{v}(\br) = \frac{1}{N_s}\sum_\bk e^{i\bk\cdot\br} v(\bk).
\end{equation}
Combining Eq.~\eqref{eq: def. of WTO} and Eq.~\eqref{eq: def. of u and v in real space}, it is clear that $\Gamma^\dagger_\bR$ is \textit{compactly supported} in real space if and only if $u(\bk)$ and $v(\bk)$ are both TPs with a \textit{finite} degree. Denoting the degree of these TPs as $(d_x,d_y)$, $\Gamma^\dagger_\bR$ is then compactly supported within a spatial region centered at $\bR$ of size $(2d_x+1)\times (2d_y+1)$. Notice that, for our purpose, $\Gamma^\dagger_\bR$ and $\Gamma_{\bR'}$ need not anticommute for general $\bR \neq \bR'$ (unless their supports are disjoint), thus they are not true Wannier orbitals and hence the name ``Wannier-type" \cite{Dubail_Read, Read2017CompactWannier}, just to emphasize their nature as an occupied orbital that is localized in real space.

Since a Gaussian state is determined by the ratio $M^p(\bk)=v(\bk)/u(\bk)$, the WTO centered at $\bR$ is clearly not uniquely defined, as multiplying $v(\bk)$ and $u(\bk)$ by a common factor $f(\bk)$ would leave $M^p(\bk)$ invariant, while by Eq.~\eqref{eq: def. of WTO} this would generally change the shape of the WTO. Nevertheless, if $M^p(\bk)$ is a ratio of TPs of finite degrees, we can show that there exists a \textit{maximally localized} WTO, which is unique up to lattice translations.

\subsubsection{Unique maximally localized Wannier-type orbital}

Let us recast $M^p(\bk)=v(\bk)/u(\bk)$ as a ratio of two \textit{regular} polynomials of $e^{ik_x}$ and $e^{ik_y}$ \footnote{By \textit{regular} polynomials of $e^{ik_x}$ and $e^{ik_y}$, we mean the class of polynomials where only non-negative powers of variables are allowed. This is contrasted with the Laurent polynomials as discussed around Eq.~\eqref{eq: general TPs}.}:
\begin{equation}
    M^p(\bk)= \frac{V(\bk)}{U(\bk)},\;\text{with}\;\; U(\bk),V(\bk) \in \mathbb{C}[e^{ik_x}, e^{ik_y}]
\end{equation}
This conversion can always be done: if $u(\bk)$ and $v(\bk)$ are Laurent polynomials of $e^{i k_{x}}$ and $e^{i k_{y}}$ with degree $(d_x,d_y)$, then multiplying both by $e^{i(d_x k_x+d_yk_y)}$ produce $U(\bk)$ and $V(\bk)$, which are regular polynomials with degree ($2d_x,2d_y$). Note that such a conversion from $u(\bk)$ and $v(\bk)$ to $U(\bk)$ and $V(\bk)$ only translate the position of the WTO without affecting its shape (c.f. Eqs.~\eqref{eq: def. of WTO} and~\eqref{eq: def. of u and v in real space}).   Furthermore, there must exist two \textit{coprime} regular polynomials  $U_*(\bk),V_*(\bk) \in \mathbb{C}[e^{ik_x}, e^{ik_y}]$ (namely there are no common non-unit polynomial factors among them) whose ratio equals to $M^p(\bk)$, hence
\begin{equation}\label{eq: UFD}
    V_*(\bk) U(\bk) = V(\bk)U_*(\bk) 
    \implies \begin{cases} 
    U(\bk) = F(\bk) U_*(\bk),\\  V(\bk) = F(\bk) V_*(\bk)
    \end{cases}
\end{equation}
where $F(\bk)\in \mathbb{C}[e^{ik_x}, e^{ik_y}]$ is a regular polynomial of $e^{ik_x}$ and $e^{ik_y}$. In the above implication we have used the fact that the regular polynomial ring $\mathbb{C}[e^{ik_x}, e^{ik_y}]$ is a unique factorization domain (UFD), 
thus if $U_*$ and $V_*$ are coprime, they are unique (up to multiplication by a constant), and $U$ ($V$) must be proportional to $U_*$ ($V_*$) in the sense of polynomial ring. It is obvious from Eqs.~\eqref{eq: def. of WTO} and~\eqref{eq: def. of u and v in real space} that the multiplication by $F(\bk)\in \mathbb{C}[e^{ik_x}, e^{ik_y}]$ cannot decrease the size of support of the WTO \textit{in any direction}. We thus conclude that the coprime polynomials $U_*(\bk),V_*(\bk) \in \mathbb{C}[e^{ik_x}, e^{ik_y}]$ are associated to the \textbf{maximally localized} Wannier-type orbital. 

Denoting the degree of the regular polynomials $U_*(\bk)$ and $V_*(\bk)$ as $(\mathcal{M}_x,\mathcal{M}_y)$, the maximally localized WTO has a support of size $(\mathcal{M}_x+1)\times (\mathcal{M}_y+1)$, and any WTO of size $\mathcal{W}_x\times \mathcal{W}_y$ must obey $\mathcal{W}_{x,y}\geq \mathcal{M}_{x,y}+1$. 

\subsection{Entanglement properties}\label{subsec: EE, BCS}

Here we discuss entanglement properties of GfTNS to motivate the connection between \ref{itm: P4} and \ref{itm: P1}. Tensor network states constructions provide many-body states that obey an ``area-law" entanglement, which capture a large class of ground states of local Hamiltonian. The intuitive relationship between TNS and ``area-law" entanglement can be understood from the fact that entanglement of the physical state arises from that of the virtual state with only nearest-neighbor correlations, see Eqs.~\eqref{eq: reference virtual state} and~\eqref{eq: real-space M}. To be more concrete, consider the Schmidt decomposition of the physical state
\begin{equation}\label{eq: def of Schmidt}
	\ket{\psi_p} = \sum_{\alpha=1}^{D_A} \lambda_\alpha \ket{\psi_{A,\alpha}}\otimes \ket{\psi_{\bar{A},\alpha}},
\end{equation}
with $0< \lambda_a \leq 1$. By construction, $\{\ket{\psi_{A,\alpha}}\}$ and $\{\ket{\psi_{\bar{A},\alpha}}\}$ are orthonormal sets of many-body states belonging to the Hilbert spaces of region $A$ and its complement $\bar{A}$, respectively. In this work we focus on the the R{\'e}nyi-0 entanglement entropy, $S_0(A)$, which is defined by the rank of the reduced density matrix $\rho_A = \Tr_{\bar{A}}[\ket{\psi_p}\bra{\psi_p}]$:
\begin{equation}
 S_0(A) : = \log_2 (D_A).
\end{equation}
The area-law entanglement, as specified in \ref{itm: P1}, states that the upper bound of $S_0(A)$ is given by the size of the partition boundary $\partial A$. Below, we explain how the structure of GfTNS as explained in Sec. \ref{subsec: GfTNS formalism} (and specified by \ref{itm: P4}), implies the area-law entanglement stated in \ref{itm: P1}.

Notice that, before the virtual-state projection, the entire Gaussian fiducial state $\ket{\psi_{vp}}$ is local in position and hence admits the decomposition $\ket{\psi_{vp}}  = \ket{\psi_{vp}}_A \otimes \ket{\psi_{vp}}_{\bar{A}}$. On the other hand, the virtual state can be decomposed as $\ket{\psi_{v}}  = \ket{\psi_{v}}_A \otimes \ket{\psi_{v}}_{\bar{A}} \otimes \ket{\psi_v}_{\partial A}$. Here $\ket{\psi_{vp}}_A$ and $\ket{\psi_{v}}_A$ belong to the Hilbert space $\mathcal{H}_A$ where local degrees of freedom (both physical and virtual) are completely supported within region $A$, while $\ket{\psi_v}_{\partial A}$ is composed of entangled pair states whose bonds cross the boundary $\partial A$. To understand the Schmidt decomposition of the physical state $\ket{\psi_p}=\braket{\psi_v|\psi_{vp}}$, it is useful to consider the following two linear maps:
\begin{equation}
\begin{split}
    T_A :={}_A\braket{\psi_v|\psi_{vp}}_A: \mathcal{H}^{(v)}_{\partial A, A} \rightarrow \mathcal{H}^{(p)}_A \\
    T_{\bar{A}} :={}_{\bar{A}}\braket{\psi_v|\psi_{vp}}_{\bar{A}}: \mathcal{H}^{(v)}_{\partial A, \bar{A}} \rightarrow \mathcal{H}^{(p)}_{\bar{A}}.
\end{split}
\end{equation}
In constructing $T_A$, we have taken the fiducial state $\ket{\psi_{vp}}_A$ within $\mathcal{H}_A$ and performed all the virtual state projection internal to region $A$. The resulting object has unspecified virtual legs on the boundary $\partial A$, as well as unspecified physical legs within $A$, and hence can be interpreted as a linear map from the Hilbert space $\mathcal{H}^{(v)}_{\partial A, A}$ of virtual modes on the boundary $\partial A$ (and within $A$), to the Hilbert space $\mathcal{H}^{(p)}_{A}$ of physical modes supported within $A$. The construction for $T_{\bar{A}}$ is similar. The physical state $\ket{\psi_p} \in \mathcal{H}^{(p)}_A \otimes\mathcal{H}^{p}_{\bar{A}}$ can then be interpreted as acting the linear map $T_A \otimes T_{\bar{A}}$ on $\ket{\psi_v}_{\partial A}$. As is clear from Eqs. \eqref{eq: reference virtual state} and \eqref{eq: real-space M}, $\ket{\psi_v}_{\partial A}$ admits the Schmidt decomposition
\begin{equation}
    \ket{\psi_v}_{\partial A} = \sum_{\bs\in \{0,1\}^{N_{\partial A}}}\ket{\bs}_{\partial A, A}\otimes\ket{\bs}_{\partial A, \bar{A}}
\end{equation}
where $\bs$ represents a string of occupation numbers ($0$ or $1$) of length $N_{\partial A}$, defined as the number of entangled pair states that cross the boundary $\partial A$. Subsequently, 
\begin{equation}
    \ket{\psi_p} = \sum_{\bs\in \{0,1\}^{N_{\partial A}}} T_A\ket{\bs}_{\partial A, A}\otimes T_{\bar{A}}\ket{\bs}_{\partial A, \bar{A}}.
\end{equation}
While the above is not guaranteed to be a Schmidt decomposition, as the $2^{N_{\partial A}}$ terms in the above decomposition may not be mutually orthogonal, it nevertheless provides an upper bound on the Schmidt rank. Hence, for a TNS specified by \ref{itm: P4} and with a rectangular region $A$ with $l_x \times l_y$ sites, we have
\begin{equation}
	S_0(A) \leq N_{\partial A} =  2n_x l_y +2n_y l_x.
\end{equation}
This proves \ref{itm: P4} $\implies$ \ref{itm: P1}. 

The subsequent focus of this work is to show \ref{itm: P1} $\implies$ \ref{itm: P4} for generic 2D fermionic Gaussian states of the form in Eq.~\eqref{eq: BCS Gaussian form}. The equivalence between \ref{itm: P1} and \ref{itm: P4} is well-known for 1D MPS. 
This success in 1D strongly motivates our consideration in 2D, but the corresponding proof is substantially more challenging. In this work, we shall thus consolidate this connection for generic 2D fermionic Gaussian states by establishing intermediate relations with other physical properties of GfTNS, which are the content of \ref{itm: P2} and \ref{itm: P3}. 

\section{Expressivity without $U(1)$ symmetry}\label{sec: single-band BCS}

To show the equivalence of the four properties of BCS GfTNSs introduced in Sec. \ref{sec:intro}, we establish proofs in the following order:  \ref{itm: P4} $\implies$ \ref{itm: P1} $\implies$ \ref{itm: P2} $\implies$ \ref{itm: P3} $\implies$ \ref{itm: P4}. Notice that our discussion in the Sec.~\ref{subsec: EE, BCS} already proves \ref{itm: P4} $\implies$ \ref{itm: P1}. We thus move on to the first non-trivial relation.

\subsection{\ref{itm: P1} $\implies$ \ref{itm: P2}}
Consider a region $A$ of size $(2n_x+1)\times l_y$, see Fig.~\ref{fig: GfTNS structure}(b), which contains $(2n_x+1)l_y =N_A$ number of physical fermionic modes. Assuming \ref{itm: P1}, we have 
\begin{equation}
    S_0(A) \leq  N_A + 2n_y(2n_x+1) - l_y,
\end{equation}
and hence by choosing $l_y>2n_y(2n_x+1)$, we ensure that $S_0(A) < N_A$. Importantly, for a many-body Gaussian state, the property $S_0(A) < N_A$ implies the existence of a Wannier-type orbital that is compactly supported within region $A$. To see this clearly, recall that the reduced density matrix $\rho_A$ of a Gaussian state is Gaussian, i.e., $\rho_A \propto \exp(\mathcal{K}_A)$ where the so-called entanglement Hamiltonian $\mathcal{K}_A$ is quadratic in terms of the fermionic operators $\{b_{i\in A},b^\dagger_{i \in A}\}$. Then there must exist a linear Bogoliubov transformation of $\{b_{i\in A},b^\dagger_{i \in A}\}$ that  diagonalizes $\mathcal{K}_A$, such that
\begin{equation}
	\rho_A = \otimes_{\alpha=1}^{N_A} \left[p_\alpha \ket{0_\alpha}\bra{0_\alpha}+(1-p_\alpha)\ket{1_\alpha}\bra{1_\alpha}\right]
\end{equation}
with $0\leq p_\alpha \leq 1$, and $N_A$ is the number of physical modes inside region $A$. If $S_0(A) < N_A$, there must exist at least one state with either $p_\alpha=0$ or $p_\alpha=1$, which implies the existence of a WTO (c.f. Eq.~\eqref{eq: def. of WTO}) that is compactly supported within region $A$ and annihilates the Gaussian physical state $\ket{\psi_p}$. 

Following our discussion in Sec.~\ref{subsec: WTO, BCS}, regarding the existence and uniqueness of the \textit{maximally localized} WTO, we conclude that $2n_x \geq \mathcal{M}_x$, where $\mathcal{M}_a+1$ is the size of the \textit{maximally localized} Wannier-type orbital along the $a$-direction. Repeating the above analysis by considering a region of size $l_x \times (2n_y+1)$ with $l_x > 2n_x(2n_y+1)$, we conclude that $2n_y \geq \mathcal{M}_y$. As there exists a maximally localized WTO with support $(\mathcal{M}_x+1)\times (\mathcal{M}_y+1)$, there must also exist a compactly supported Wannier-type orbital of size $(2n_x+1)\times(2n_y+1)$ (depicted as the grey shaded region in Fig.~\ref{fig: GfTNS structure}(b)). Hence \ref{itm: P1} implies by \ref{itm: P2}.

\subsection{\ref{itm: P2} $\implies$ \ref{itm: P3}}
Given a compactly supported Wannier-type orbital (WTO), $\Gamma^\dagger_\bR =\sum_{\br}\left(\tilde{u}(\br) b_{\bR+\br} -\tilde{v}(\br)b^\dagger_{\bR+\br}\right)$, we consider its Fourier transform:
\begin{equation}
\begin{split}
\gamma^\dagger_\bk & = \left(\sum_\br e^{-i\bk\cdot\br}\tilde{u}(\br)\right) b_{-\bk} -\left(\sum_\br e^{-i\bk\cdot\br}\tilde{v}(\br)\right) b^\dagger_{\bk} \\
    & = u(\bk)b_{-\bk}-v(\bk)b^\dagger_{\bk}
\end{split}
\end{equation}
Assuming \ref{itm: P2}, we have $\tilde{u}(\br) = \tilde{v}(\br) = 0$ for $\abs{\br_{x,y}} > n_{x,y}$, hence  $u(\bk)=\sum_\br e^{-i\bk\cdot\br}\tilde{u}(\br)$ and $v(\bk)=\sum_\br e^{-i\bk\cdot\br}\tilde{v}(\br)$ are Laurent polynomials of $(e^{ik_x}, e^{ik_y})$ with degree $(n_x,n_y)$. Since $\Gamma^\dagger_\bR \ket{\psi_p}=0\;\;\forall \bR$ (by virtue of translation symmetry), we have $\gamma^\dagger_\bk \ket{\psi_p}=0\;\;\forall \bk$, i.e., 
\begin{equation}
\begin{split}
&(-v(\bk)b^\dagger_\bk+u(\bk)M^p(\bk)b_{-\bk}b^\dagger_{-\bk} b^\dagger_{\bk})\ket{0}_p=0 \\ &\quad\quad\quad\implies M^p(\bk)=\frac{v(\bk)}{u(\bk)}. 
\end{split}
\end{equation}
Notice that at this stage $M^p(\bk)$ is not necessarily of the form stated in \ref{itm: P3} 
because the symmetry properties of $u(\bk)$ and $v(\bk)$ are not yet clear at this stage.

To fill in this gap, we again make use of the unique maximally localized Wannier-type orbital. Given \ref{itm: P2}, the support of the maximally localized WTO, $(2\mathcal{M}_x+1)\times(2\mathcal{M}_y+1)$, would obey $\mathcal{M}_{x,y}\leq n_{x,y}$. Following the discussion in Sec. \ref{subsec: WTO, BCS}, making use of the fact that the polynomial ring $\mathbb{C}[e^{ik_x}, e^{ik_y}]$ is a unique factorization domain (UFD), we know that $M^p(\bk)$ admits a representation as a ratio of two \textit{coprime} regular polynomials, i.e., $M^p(\bk)
= V(z_x,z_y)/U(z_x,z_y)$ with $z_{x,y}= e^{ik_{x,y}}$, and the regular polynomial degrees satisfy $\deg U,\deg V\le (2\mathcal{M}_x,2\mathcal{M}_y)$. The fermionic antisymmetry $M^p(-\bk)=-M^p(\bk)$ is equivalent to $M^p(z_x^{-1},z_y^{-1})=-M^p(z_x,z_y)$. Defining a ``reversed" pair of regular polynomials:
\begin{equation}\label{eq: reversing polynomials}
\begin{split}
    U^\sharp(z_x,z_y)
=
z_x^{2\mathcal{M}_x}z_y^{2\mathcal{M}_y}
U(z_x^{-1},z_y^{-1}),\\
V^\sharp(z_x,z_y)
=
z_x^{2\mathcal{M}_x}z_y^{2\mathcal{M}_y}
V(z_x^{-1},z_y^{-1}).
\end{split}
\end{equation}
we see that $M^p(z_x,z_y)=-V^\sharp(z_x,z_y)/U^\sharp(z_x,z_y)$, which gives another representation of $M^p(z_x,z_y)$ as a ratio of two coprime polynomials. By the property of UFD, we must have $U^\sharp=cU$ and $V^\sharp=-cV$ for some constant $c\in\mathbb C^\times$. Notice that $c=\pm 1$, because reversing the polynomial twice according to Eq.~\eqref{eq: reversing polynomials} is an identity operation. The corresponding Laurent polynomials, $u(\bk)=z_x^{-\mathcal{M}_x}z_y^{-\mathcal{M}_y}U(z_x,z_y)$ and $v(\bk)=z_x^{-\mathcal{M}_x}z_y^{-\mathcal{M}_y}V(z_x,z_y)$, thus satisfy
\begin{equation}
    u(-\bk)=\pm \,u(\bk), \;\text{ and }\;\;
v(-\bk)=\mp\,v(\bk).
\end{equation}
Notice that the degree of the above Laurent TPs are $(\mathcal{M}_x,\mathcal{M}_y)\leq (n_x,n_y)$, and hence with the symmetry properties demonstrated above we have established the entire content of \ref{itm: P3}. Below, we always denote $u(\bk)$ as a parity-even TP and $v(\bk)$ as a parity-odd TP, and hence either $M^p(\bk)= v(\bk)/u(\bk)$ or $M^p(\bk)= u(\bk)/v(\bk)$. 

\subsection{\ref{itm: P3} $\implies$ \ref{itm: P4}}\label{sec: P3 implies P4}
We now turn to the more challenging part. As we have seen in Sec.~\ref{sec: GfTNS review} by counting,  the number of independent parameters required to specify $M^p(\bk)$ of the form in \ref{itm: P3}, $\mathcal{N}=4n_xn_y+2n_x+2n_y$, equals to the number of independent parameters possessed by a 2D GfTNS with horizontal and vertical bond dimensions $2^{n_x}$ and $2^{n_y}$, respectively. It is therefore well-motivated to consider whether a 2D GfTNS with bond dimensions $2^{n_x}$ and  any $2^{n_y}$ can realize \textit{all} $M^p(\bk)$ of the corresponding degrees (i.e., \ref{itm: P3}). Recall \ref{itm: P3} includes two possibilities: either $M^p(\bk)= v(\bk)/u(\bk)$ or $M^p(\bk)= u(\bk)/v(\bk)$. Below we first assume $M^p(\bk)= v(\bk)/u(\bk)$ and prove \ref{itm: P3} $\implies$ \ref{itm: P4} in this case. The other possibility, with $M^p(\bk)= u(\bk)/v(\bk)$, can then be treated easily by a particle-hole transformation, as explained at the end in Sec.~\ref{sec: Concluding remarks for P3 implies P4}.

To rigorously establish this connection, we recall from Sec.~\ref{sec: GfTNS review} the relation between the momentum-space wavefunction and the tensor network parameters:
\begin{equation}\label{eq: relation between the momentum-space wavefunction and the tensor network parameters, np=1}
    M^p(\bk) = N^{pv}[N^{vv}+M(\bk)]^{-1}(N^{pv})^T,
\end{equation}
where we have set $N^{pp}=0$ as we focus on single-band BCS states with only one physical fermion per tensor. Our goal in this section is to show that Eq.~\eqref{eq: relation between the momentum-space wavefunction and the tensor network parameters, np=1} can represent any generic $M^p(\bk)$ as a ratio of TPs in the form specified by \ref{itm: P3}. To help readers understand our proof for the 2D TNS, which is technically involved, let us first prove the analogous relation for 1D TNS. The simpler approach developed for 1D TNS can be extended to 2D TNS, which we describe in Sec.~\ref{subsec: 2D tuning u and v} and Sec.~\ref{subsec: Chevalley}. Further details are provided in Appendix~\ref{appendix: Lagrange}.

\subsubsection{Expressivity of 1D TNS}\label{subsec: 1D TNS}
Recall the counting argument discussed in Sec.~\ref{subsec: GfTNS formalism}, for 1D TNS with $n$ virtual fermion modes on each leg of the tensor, there are $2n$ independent TNS parameters. This motivates us to restrict to the following form of the antisymmetric virtual mode coupling matrix:
\begin{subequations}\label{eq: Nvv restricted form in 1D}
\begin{align}
    N^{vv} &= \begin{pmatrix}
        \mathbf{0}_n & N^{vv}_{LR} \\ -N^{vv}_{LR} & \mathbf{0}_n
    \end{pmatrix},\\
    N^{vv}_{LR} &=\text{diag}([N^{vv}_{LR}]_1,\; ...\;,\; [N^{vv}_{LR}]_n),
\end{align}
\end{subequations}
which possesses $n$ independent complex parameters. Substituting this and Eq.~\eqref{eq: Mk_1D} into Eq.~\eqref{eq: relation between the momentum-space wavefunction and the tensor network parameters, np=1}, we obtain
\begin{equation}\label{eq: 1D_Mp(k) from TNS}
\begin{split}
   M^p(k) = \sum_{\lambda=1}^n \frac{N^{pv}_\lambda N^{pv}_{n+\lambda}(e^{ik}-e^{-ik})}{(e^{ik}-[N^{vv}_{LR}]_\lambda)(e^{-ik}-[N^{vv}_{LR}]_\lambda)}
\end{split}
\end{equation}
from which it is clear that the other $n$ TNS parameters relevant for constructing $M^p(k)$ are $\{N^{pv}_\lambda N^{pv}_{n+\lambda}\vert \lambda=1,...,n\}$. The common denominator of the right hand side of Eq.~\eqref{eq: 1D_Mp(k) from TNS} is simply 
\begin{equation}\label{eq: det[Nvv+M] in 1D}
    \det[N^{vv}+M(k)] = \prod_{\lambda=1}^n (e^{ik}-[N^{vv}_{LR}]_\lambda)([N^{vv}_{LR}]_\lambda-e^{-ik})
\end{equation}
which is even under $k\rightarrow - k$, and hence we identify it as the parity-even TP $u(k)$ in \ref{itm: P3}. Notice that every parity-even TP $u(k)$ of degree $n$ can be equivalently viewed as a degree-$n$ Laurent polynomial of a complex variable $z$ (with the identification $z:=e^{ik}$) that is symmetric under $z \leftrightarrow 1/z$. By the fundamental theorem of algebra, this symmetric polynomial is uniquely determined (up to a proportionality factor) by its $n$ roots $\{[N^{vv}_{LR}]_{\lambda=1,...,n}\}$ (together with their symmetric partners $\{1/[N^{vv}_{LR}]_{\lambda=1,...,n}\}$). Hence the form in Eq.~\eqref{eq: det[Nvv+M] in 1D} implies that, upon tuning the TNS parameters $\{[N^{vv}_{LR}]_{\lambda=1,...,n}\}$, we can represent any parity-even TP $u(k)$ using $\det[N^{vv}+M(k)]$, up to a proportionality constant.

Next, we identify $v(k)= \det[N^{vv}+M(k)] M^{p}(k)$ as the parity-odd TP in \ref{itm: P3}, which can be recast as
\begin{equation}\label{eq: v(k) in 1D}
\begin{split}
    v(k) = \prod_{\lambda=1}^n &(e^{ik}-[N^{vv}_{LR}]_\lambda)([N^{vv}_{LR}]_\lambda-e^{-ik}) \\
    &\times
    \left[ \sum_{\lambda'=1}^n \frac{-N^{pv}_{\lambda'} N^{pv}_{n+\lambda'}}{e^{ik}-[N^{vv}_{LR}]_{\lambda'}} - (k \rightarrow - k)\right].
\end{split}
\end{equation}
Using the method of Lagrange interpolation, we can see that the above form can represent any parity-odd TP of degree $n$:
\begin{equation}
    v(k) = \sum_{a=1}^n v_a \sin(ak),\;\;v_a\in\mathbb{C}.
\end{equation}
Consider the following set of $n$ interpolation points: $\{e^{ik_b} = [N^{vv}_{LR}]_b \,\vert\, b=1,...,n\}$, we notice that $v(k_\lambda)$ can be independently tuned to an arbitrary complex value by varying $N^{pv}_{\lambda}N^{pv}_{n+\lambda}$. Hence, the $n$ TP coefficients 
\begin{equation}
    v_a = \sum_{b=1}^n
[\mathcal{S}^{-1}]_{ab}v(k_b),\;\mathcal{S}_{ab} \equiv \frac{([N^{vv}_{LR}]_b)^a-([N^{vv}_{LR}]_b)^{-a}}{2i}
\end{equation}
can also attain arbitrary complex values upon tuning the $n$ independent TNS parameters given by $\{N^{pv}_\lambda N^{pv}_{n+\lambda}\vert \lambda=1,...,n\}$. 

This finishes the proof showing that a 1D TNS with bond dimension $2^n$ can generically represent any  BCS Gaussian state (Eq.~\eqref{eq: BCS Gaussian form}) whose momentum-space wavefunction $M^p(k)$ is a ratio of two TPs with degree $n$. Next, we follow a similar strategy to prove \ref{itm: P3} $\implies$  \ref{itm: P4} for 2D TNS. 

\subsubsection{Tuning $u(\bk)$ and $v(\bk)$ with 2D TNS}\label{subsec: 2D tuning u and v}
Similar to the above 1D analysis, in 2D we can also relate the parity-even TP $u(\bk)$ to $\det[N^{vv}+M(\bk)]$ and identify $v(\bk) = \det[N^{vv}+M(\bk)] M^p(\bk)$.
For 2D TNS, the full expression for the right-hand-side of Eq.~\eqref{eq: relation between the momentum-space wavefunction and the tensor network parameters, np=1} in terms of the $\mathcal{N}=4n_xn_y+2n_x+2n_y$ independent TNS parameters is significantly more complicated than the 1D counterpart analyzed above. We will thus proceed in two steps. First, we focus on a local region of the TNS parameter space (to be specified below) and show that in this region the TP coefficients of $u(\bk)$ and $v(\bk)$, see Eq.~\eqref{eq: general TPs}, can be independently tuned. Then in Sec.~\ref{subsec: Chevalley}, we make use of a theorem in elementary algebraic geometry (applied to the complex polynomial ring) to show that these TP coefficients can attain generic complex values by accessing the full TNS parameter space. 

Making use of gauge redundancies, we restrict to the following form for the 2D virtual mode coupling matrix:
\begin{equation}\label{eq: restricted form of N^vv}
N^{vv} = \begin{pmatrix}
  \mathbf{0}_{n_x} & N^{vv}_{LR} & N^{vv}_{LD} & N^{vv}_{LU} \\
  -N^{vv}_{LR} & \mathbf{0}_{n_x} & N^{vv}_{RD} & N^{vv}_{RU} \\
  -(N^{vv}_{LD})^T & - (N^{vv}_{RD})^T &\mathbf{0}_{n_y} & N^{vv}_{DU} \\
  -(N^{vv}_{LU})^T & -(N^{vv}_{RU})^T & -N^{vv}_{DU} & \mathbf{0}_{n_y}
\end{pmatrix},
\end{equation}
where
\begin{subequations}
\begin{align}
    N^{vv}_{LR}  &= \text{diag}([N^{vv}_{LR}]_1,\;...\;,\;[N^{vv}_{LR}]_{n_x}),\\
    N^{vv}_{DU}  &= \text{diag}([N^{vv}_{DU}]_1,\;...\;,\;[N^{vv}_{DU}]_{n_y}),
\end{align}
\end{subequations}
and $\{N^{vv}_{LD}, N^{vv}_{LU}, N^{vv}_{RD}, N^{vv}_{RU}\}$ are generic $n_x \times n_y$ matrices. Altogether, $N^{vv}$ of this form contains $4n_xn_y+n_x+n_y$ independent parameters. Motivated by the above 1D analysis, we further assert the following form for the $1\times (2n_x+2n_y)$ matrix $N^{pv}$:
\begin{equation}\label{eq: restricted form of N^{pv}}
    N^{pv} = \begin{pNiceArray}[first-row]{c|c|c|c}
      n_x & n_x & n_y & n_y\\
  N^{pv}_{L} & 1\;...\;1 & N^{pv}_{D} & 1\;...\;1 
\end{pNiceArray}, 
\end{equation}
where the size of each sub-block is indicated. As such, $N^{pv}$ contributes the remaining $n_x+n_y$ TNS parameters.

Let us now focus on a specific region in the TNS parameter space where all the $n_x \times n_y$ matrices $\{N^{vv}_{LD}, N^{vv}_{LU}, N^{vv}_{RD}, N^{vv}_{RU}\}$ are small (i.e., with all matrix elements much smaller than 1). Plugging in Eq.~\eqref{eq: Mk_2D} and expanding to the quadratic order in $\{N^{vv}_{LD}, N^{vv}_{LU}, N^{vv}_{RD}, N^{vv}_{RU}\}$, we obtain
\begin{widetext}
    \begin{equation}\label{eq: det(N^vv+M) quadratic in B}
    \begin{split}
        &\det[N^{vv}+M(\bk)] = \prod_{\lambda=1}^{n_x}\prod_{\mu=1}^{n_y}\; (e^{ik_x}-[N^{vv}_{LR}]_\lambda)(e^{-ik_x}-[N^{vv}_{LR}]_\lambda) (e^{ik_y}-[N^{vv}_{DU}]_\mu)(e^{-ik_y}-[N^{vv}_{DU}]_\mu) \\
        &\quad\quad\times \Bigg\{1+ \sum_{\lambda'=1}^{n_x}\sum_{\mu' = 1}^{n_y} \left[\frac{[N^{vv}_{LU}]_{\lambda'\mu'} [N^{vv}_{RD}]_{\lambda'\mu'}}{(e^{ik_x}-[N^{vv}_{LR}]_{\lambda'})(e^{ik_y}-[N^{vv}_{DU}]_{\mu'})}-\frac{[N^{vv}_{LD}]_{\lambda'\mu'} [N^{vv}_{RU}]_{\lambda'\mu'}}{(e^{ik_x}-[N^{vv}_{LR}]_{\lambda'})(e^{-ik_y}-[N^{vv}_{DU}]_{\mu'})}+ (\bk\rightarrow -\bk) \right]\Bigg\}.
    \end{split}
\end{equation}
\end{widetext}
As for $v(\bk)=\det[N^{vv}+M(\bk)]M^p(\bk)$, up to  linear order in $\{N^{vv}_{LD}, N^{vv}_{LU}, N^{vv}_{RD}, N^{vv}_{RU}\}$, we find

\begin{widetext}
\begin{equation}\label{eq: full v(k) to linear order in B}
\begin{split}
    v(\bk) = &\prod_{\lambda=1}^{n_x}\prod_{\mu=1}^{n_y}\;(e^{ik_x}-[N^{vv}_{LR}]_\lambda)(e^{-ik_x}-[N^{vv}_{LR}]_\lambda) (e^{ik_y}-[N^{vv}_{DU}]_\mu)(e^{-ik_y}-[N^{vv}_{DU}]_\mu) \\
    &\times\Big[ \sum_{\lambda'=1}^{n_x}\frac{-[N^{pv}_L]_{\lambda'}}{e^{ik_x}-[N^{vv}_{LR}]_{\lambda'}} + \sum_{\mu'=1}^{n_y}\frac{-[N^{pv}_D]_{\mu'}}{e^{ik_y}-[N^{vv}_{DU}]_{\mu'}} + \sum_{\lambda'=1}^{n_x}\sum_{\mu'=1}^{n_y}\frac{ [N^{pv}_L]_{\lambda'} [N^{vv}_{RD}]_{\lambda' \mu'}-[N^{pv}_{D}]_{\mu'}[N^{vv}_{LU}]_{\lambda'\mu'}}{(e^{ik_x}-[N^{vv}_{LR}]_{\lambda'})(e^{ik_y}-[N^{vv}_{DU}]_{\mu'})} \\
    &\quad\quad +\;\sum_{\lambda'=1}^{n_x}\sum_{\mu'=1}^{n_y}\frac{[N^{pv}_L]_{\lambda'}[N^{vv}_{RU}]_{\lambda'\mu'}[N^{pv}_D]_{\mu'}-[N^{vv}_{LD}]_{\lambda'\mu'}}{(e^{-ik_x}-[N^{vv}_{LR}]_{\lambda'})(e^{ik_y}-[N^{vv}_{DU}]_{\mu'})} - \;(\bk\rightarrow - \bk)\;\Big].
\end{split}
\end{equation}
\end{widetext}
We remark that, despite the relatively complicated forms of Eq.~\eqref{eq: det(N^vv+M) quadratic in B} and Eq.~\eqref{eq: full v(k) to linear order in B}, they are directly analogous to the 1D expression in Eq.~\eqref{eq: v(k) in 1D}, and hence we can similarly apply the method of Lagrange interpolation as detailed there. Paritcularly, we can argue that Eq.~\eqref{eq: det(N^vv+M) quadratic in B} does represent a parity-even TP $u(\bk)$ of degree $(n_x,n_y)$, whose $\mathcal{N}/2=2n_xn_y+n_x+n_y$ TP coefficients (modulo a proportionality constant) can be independently tuned by $\mathcal{N}/2$ TNS parameters. Similarly, it can be argued that the remaining $\mathcal{N}/2$ TNS parameters can independently tune the $\mathcal{N}/2$ TP coefficients of the parity-odd TP $v(\bk)$ of degree $(n_x,n_y)$, as in Eq.~\eqref{eq: full v(k) to linear order in B}. TABLE ~\ref{table: TNS parameters without U(1)} provides a summary of this result, showing the explicit list of TNS parameters involved in tuning the TP coefficients. 
The derivation of Eq.~\eqref{eq: det(N^vv+M) quadratic in B} and Eq.~\eqref{eq: full v(k) to linear order in B} and details of Lagrange interpolation is provided in Appendix~\ref{subsec: Lagrange w/o U(1)}.

\begin{table}[t!]
\caption{List of 2D tensor network state (TNS) parameters used to tune the parity-even and parity-odd trigonometric polynomials (TPs), $u(\bk)$ and $v(\bk)$, respectively, based on Eq.~\eqref{eq: det(N^vv+M) quadratic in B} and Eq.~\eqref{eq: full v(k) to linear order in B}. Given gauge redundancies of TNS, we have restricted the virtual-mode couplings as encoded by $N^{vv}$ in the form of Eq.~\eqref{eq: restricted form of N^vv}, and the physical-virtual couplings $N^{pv}$ in Eq.~\eqref{eq: restricted form of N^{pv}}. Index $\lambda$ runs from $1$ to $n_x$, and index $\mu$ runs from $1$ to $n_y$. }\label{table: TNS parameters without U(1)}
\renewcommand{\arraystretch}{1.5} 

\begin{ruledtabular}\label{table}
\begin{center}
\begin{tabular}{c|c|c}
\text{\small TP}& $u(\bk)$ & $v(\bk)$ \\ \hline
 & $[N^{vv}_{LR}]_{\lambda}, \; [N^{vv}_{DU}]_{\mu},$  & $[N^{pv}_L]_\lambda,\;[N^{pv}_D]_\mu,$ \\
\text{\small TNS} & $[N^{vv}_{LU}]_{\lambda \mu}[N^{vv}_{RD}]_{\lambda \mu},$  & $[N^{pv}_L]_\lambda[N^{vv}_{RD}]_{\lambda\mu}-[N^{pv}_D]_\mu [N^{vv}_{LU}]_{\lambda\mu},$ \\
&$[N^{vv}_{LD}]_{\lambda \mu}[N^{vv}_{RU}]_{\lambda \mu}$& $[N^{pv}_L]_\lambda [N^{vv}_{RU}]_{\lambda\mu}[N^{pv}_D]_\mu -[N^{vv}_{LD}]_{\lambda\mu}$
\end{tabular}
\end{center}
\end{ruledtabular}
\end{table}

\subsubsection{Expressivity of 2D TNS from Chevalley's theorem}\label{subsec: Chevalley}
In the above we have focused on a local region in the TNS parameter space, where $\{N^{vv}_{LD}, N^{vv}_{LU}, N^{vv}_{RD}, N^{vv}_{RU}\} \approx 0$, and establish that $\mathcal{N}$ distinct TNS parameters can independently tune $\mathcal{N}$ TP coefficients $\{u_\bmm, v_\bmm\vert\bmm\in\mathcal{L}(n_x,n_y)\}$. Notice that in the above we have asserted $\det[N^{vv}+M(\bk)]$ to be \textit{proportional} to a generic parity-even TP $u(\bk)$, while the latter has in total $\mathcal{N}/2+1$ TP coefficients (including $u_\mathbf{0}$, see Eq.~\eqref{eq: general TPs}). Hence, in addition to the $\mathcal{N}$ independent TNS parameters, let us include a proportionality constant $\mathcal{U}\in\mathbb{C}$ for the sake of our subsequent analysis, so that the set of TNS parameters now consists of $\mathcal{N}+1$ elements
\begin{equation}\label{eq: the extended set of TNS parameters}
\begin{split}
    \text{TNS:}\;\;\{&\mathcal{U},\;[N^{pv}_L]_\lambda,\; [N^{pv}_D]_\mu,\; [N^{vv}_{LR}]_{\lambda},\; [N^{vv}_{DU}]_{\mu}, \\
    &[N^{vv}_{LD}]_{\lambda\mu},\; [N^{vv}_{LU}]_{\lambda\mu},\; [N^{vv}_{RD}]_{\lambda\mu},\; [N^{vv}_{RU}]_{\lambda\mu}\}
\end{split}
\end{equation}
with indices $1\leq \lambda\leq n_x$ and $1\leq \mu\leq n_y$. The identifications we are proposing
\begin{equation}
\begin{split}
    u(\bk) &=\mathcal{U}\det[N^{vv}+M(\bk)], \\
    v(\bk) &=\mathcal{U}M^p(\bk)\det[N^{vv}+M(\bk)],
\end{split}
\end{equation}
then leads to a complex \textit{polynomial map} from the set of TNS parameters in Eq.~\eqref{eq: the extended set of TNS parameters} to the entire set of TP coefficients $\{u_\mathbf{0}, u_{\bmm\in\mathcal{L}}, v_{\bmm\in\mathcal{L}}\}$:
\begin{equation}\label{eq: polynomial map T}
\begin{split}
    \mathcal{T}: \mathbb{C}^{\mathcal{N}+1} \rightarrow \mathbb{C}^{\mathcal{N}+1},\;\text{TNS parameters}\;\mapsto\text{TP coefficients}
\end{split}
\end{equation}
Having analyzed the behavior of the polynomial map $\mathcal{T}$ \textit{locally} around some reference TNS, we have argued that the corresponding image of $\mathcal{T}$ \textit{locally} spans an $(\mathcal{N}+1)$-dimensional space. Using this local property, we now prove a stronger statement that the image of $\mathcal{T}$ actually covers the entire target space $\mathbb{C}^{\mathcal{N}+1}$ up to some measure-0 hypersurface, which amounts to proving the \textit{generic expressivity} of TNS. This is achieved by invoking the following theorem from algebraic geometry \cite{hartshorne2013algebraic, vakil2025rising,nlab_chevalley}:
\begin{theorem}[Chevalley's theorem]
    If $f: X \rightarrow Y$ is a polynomial map of varieties, then the image of a constructible set is constructible. In particular, the image of $f$ is constructible. 
\end{theorem}
Let us unpack a few terminologies. First, an algebraic \textit{variety} is a set of solutions of a system of polynomial equations over an algebraically closed field, which we take to be $\mathbb{C}$ here. For our application, this variety is taken to be the entire complex space $\mathbb{C}^{\mathcal{N}+1}$. Correspondingly, a \textit{constructible set} can be defined by the Zariski topology on $\mathbb{C}^{\mathcal{N}+1}$, where a \textit{closed set} is defined as a set of solutions (points) associated to some polynomial equations, and \textit{open sets} are complements of closed sets in $\mathbb{C}^{\mathcal{N}+1}$.  Hence, a closed set is either the entire space $\mathbb{C}^{\mathcal{N}+1}$ or a lower-dimensional hypersurface (i.e., with a zero Lebesgue measure in $\mathbb{C}^{\mathcal{N}+1}$).
Now we provide the definition of a constructible set \cite{hartshorne2013algebraic, vakil2025rising}:
\begin{definition}[Constructible set]
    A constructible set is a finite union of locally closed sets, i.e. it is of the form $\cup_{i=1}^n (U_i \cap V_i)$, where $U_i$ is an open set and $V_i$ is a closed set. 
\end{definition}
Applying Chevalley's theorem to the polynomial map $\mathcal{T}$ in Eq.~\eqref{eq: polynomial map T}, we see that the image of $\mathcal{T}$ is constructible, and hence a finite union of locally closed sets, $\cup_{i=1}^n (U_i \cap V_i)$. From Sec.~\ref{subsec: 2D tuning u and v}, we have seen that there exists a local region of TNS parameter space where the image of $\mathcal{T}$ spans a $(\mathcal{N}+1)$-dimensional space. This implies one of the closed sets $V_a$ must be the entire complex space $\mathbb{C}^{\mathcal{N}+1}$, and $U_a \cap V_a = U_a$ is an open set which must correspond to either the entire $\mathbb{C}^{\mathcal{N}+1}$, or $\mathbb{C}^{\mathcal{N}+1}\setminus \mathcal{P}$ with $\mathcal{P}$ a lower-dimensional hypersurface (a closed set). Therefore, the image of $\mathcal{T}$ covers the entire $\mathbb{C}^{\mathcal{N}+1}$ up to at most some lower-dimensional hypersurface with a zero Lebesgue measure.

\subsubsection{Concluding remarks}\label{sec: Concluding remarks for P3 implies P4}
Let us organize what we have established so far in Sec.~\ref{sec: P3 implies P4}. We have demonstrated, with the help of Chevalley's theorem from algebraic geometry, that the 2D GfTNS (whose structure is defined in Sec.~\ref{subsec: GfTNS formalism}) with horizontal (vertical) bond dimension $2^{n_x}$ ($2^{n_y}$) can represent any \textit{generic} BCS state in Eq.~\eqref{eq: BCS Gaussian form}, where $M^p(\bk) = v(\bk)/u(\bk)$ with $u(\bk)$ a parity-even trigonometic polynomial (TP) and $v(\bk)$ a parity-odd TP, both with TP degree $(n_x,n_y)$. 

Let us clarify what we mean by ``generic" here: from the above discussion of applying Chevalley's theorem, we have seen that the image of the polynomial map $\mathcal{T}$ (see Eq.~\eqref{eq: polynomial map T}) does not necessarily cover the entire parameter space of the TP coefficients. Therefore, there could exist fine-tuned BCS states, with fine-tuned $u(\bk)$ and $v(\bk)$, which cannot be exactly represented by the prescribed TNS structure. However, these fine-tuned states only belong to a sub-dimensional space in the entire TP parameter space. These fine-tuned states can be approximated arbitrarily well by moving only infinitesimally away from these sub-dimensional space, and landing into the region where there is an exact TNS representation (with the same fixed bond dimension) as guaranteed by Chevalley's theorem. It is in this generic sense that we establish \ref{itm: P3} $\implies$ \ref{itm: P4}.

For completeness, we also need to entertain the possibility that the physical BCS state $\ket{\psi_p}$ is described by  $M^p(\bk) = u(\bk)/v(\bk)$ (compared to what has been assumed above, i.e., $M^p(\bk) = v(\bk)/u(\bk)$), for parity-even TP $u(\bk)$ and parity-odd TP $v(\bk)$. There are two ways to proceed. One is a quick fix at the cost of increasing the number of virtual fermionic modes by one: notice that $M^p(\bk) = [u(\bk)f(\bk)]/[v(\bk)f(\bk)]$, with $f(\bk)=\sin k_x=-f(-\bk)$, and hence $u(\bk)f(\bk)$ can serve the role as a parity-odd TP and $v(\bk)f(\bk)$ serve the role as a parity-even TP, both with degree $(n_x+1,n_y)$. As such, our previous discussion applies identically to this case at the cost of only increasing the number of virtual fermion modes on some bonds by one. For practical purposes of TNS simulations, this cost is only marginal.

There is an alternative approach that does not require the addition of extra virtual modes. To that end, let us consider an auxiliary BCS state 
\begin{equation}
    \ket{\psi'_p}\propto \sideset{}{'}\prod_{\bk}(1+\widetilde{M}^p(\bk)b^\dagger_\bk b^\dagger_{-\bk}) \ket{0_p},\;\widetilde{M}^p(\bk)=\frac{v(\bk)}{u(\bk)}.
\end{equation}
Following our above discussion, $\ket{\psi'_p}$ in general has a GfTNS representation whose bond dimensions are given by the TP degree $(n_x,n_y)$. Now consider a particle-hole conjugation of this state, where $b^\dagger_\bk \overset{\text{p.h.}}{\longmapsto} b_{-\bk}$ and $\ket{0_p}\overset{\text{p.h.}}{\longmapsto} \ket{1_p}= \sideset{}{'}\prod_{\bk}b^\dagger_{\bk}b^\dagger_{-\bk}\ket{0_p}$. Correspondingly, we have
\begin{equation}
    \ket{\psi'_p} \overset{\text{p.h.}}{\longmapsto} \ket{\psi_p}.
\end{equation}
Thus with $M^p(\bk)=u(\bk)/v(\bk)$, $\ket{\psi_p}$ can be constructed from the particle-hole conjugate of a Gaussian TNS with $n_x$ $(n_y)$ number of virtual complex fermion modes per horizontal (vertical) leg. Altogether, we have completed the proof that \ref{itm: P3} $\implies$ \ref{itm: P4}.

Notice that particle-hole conjugation does not change the entanglement properties (does not change the structure of Schmidt decomposition), so either way we still have \ref{itm: P4} $\implies$ \ref{itm: P1}. Therefore, we have achieved the goal of establishing the equivalence of the four properties \ref{itm: P1}, \ref{itm: P2}, \ref{itm: P3} and \ref{itm: P4}, for BCS states. Importantly, our results show that an area law for the R{\'e}nyi-0 entropy does imply an exact Gaussian fermionic TNS representation with a finite bond dimension determined by the R{\'e}nyi-0 entropy (by \ref{itm: P1} $\iff$ \ref{itm: P4}). This is a highly non-trivial extension of the 1D MPS result. Besides, the minimal bond dimension for the TNS representation is shown to be given by the size of the maximally localized Wannier-type orbital for the BCS state (by \ref{itm: P2} $\iff$ \ref{itm: P4}), and closely related to the analytical form of the momentum-space wavefunction as a ratio of trigonometric polynomials with a finite degree. Next, we establish similar properties for $U(1)$ charge conserving band insulators with a $U(1)$-symmetric GfTNS representation.

\section{$U(1)$-symmetric Gaussian Fermion Tensor network}\label{sec: U(1) TNS review}

In this section we describe the GfTNS construction for a generic 2D band insulator with $n^{(-)}_p$ filled bands and $n^{(+)}_p$ empty bands. This sets the stage for discussing their GfTNS expressivity in connection to entanglement properties, as discussed later in Sec.~\ref{sec: U(1) insulator, expressivity}.  Such a band insulator can be generically expressed as the following Gaussian state \cite{Dubail_Read}:
\begin{equation}\label{eq: U(1)-symmetric Gaussian state in k-space}
    \ket{\psi_p} \propto \prod_\bk e^{ [M^p(\bk)]_{\bar{\beta} \beta}c^\dagger_{(\bk,\bar{\beta})} c_{(\bk,\beta)}} \vert\underbrace{11...1}_{n^{(-)}_p}\underbrace{00...0}_{n^{(+)}_p}\rangle_p,
\end{equation}
where $\ket{11...100...0}_p$ represents a reference physical state where the first $n^{(-)}_p$ orbitals (labeled by $\beta$) are occupied, and the last $n^{(+)}_p$ orbitals (labeled by $\bar{\beta}$) are empty, for every momentum $\bk$. This can be understood by noticing that Eq. \eqref{eq: U(1)-symmetric Gaussian state in k-space} implies the following Slater determinant form
\begin{equation}
    \ket{\psi_p} \propto \prod_\bk \prod_\beta\left(c^\dagger_{(\bk,\beta)}+\sum_{\bar{\beta}}[M^p(\bk)]_{\bar{\beta}\beta}c^\dagger_{(\bk,\bar{\beta})}\right)\ket{0}_p.
\end{equation}
On the other hand, a generic band insulator takes the Slater-determinant form $\prod_\bk \prod_m f^\dagger_{\bk,m}\ket{0}_p$, where $f^\dagger_{\bk,m}$ is the creation operator for the $m$-th band at momentum $\bk$, which can be related to the orbital operators by $f^\dagger_{\bk,m} = \sum_{\beta} [A(\bk)]_{\beta m}c^\dagger_{(\bk,\beta)}+\sum_{\bar{\beta}}[B(\bk)]_{\bar{\beta}m}c^\dagger_{(\bk,\bar{\beta})}$. From this, it can be easily seen that the Gaussian form in Eq. \eqref{eq: U(1)-symmetric Gaussian state in k-space} can indeed represent the Slater-determinant band insulating state given that $M^p(\bk) = B(\bk) A^{-1}(\bk)$. 

We would like to further remark that this formulation does not require a spectral gap and hence applies to gapless states with a fixed number of occupied bands as well (e.g. semimetals associated to Hamiltonians with a quadratic band touching). Here we use the word ``insulator" just to highlight the aspect of $U(1)$ charge conservation in contrast to the BCS superconductor considered in previous sections.  
With this understanding, we now detail the GfTNS construction for such states.

\subsection{Virtual-state projection and particle-hole transformation}\label{sec: U(1)-symmetric TNS construction}
\begin{figure}[t]
    \centering
    \resizebox{0.6\columnwidth}{!}{\includegraphics[]{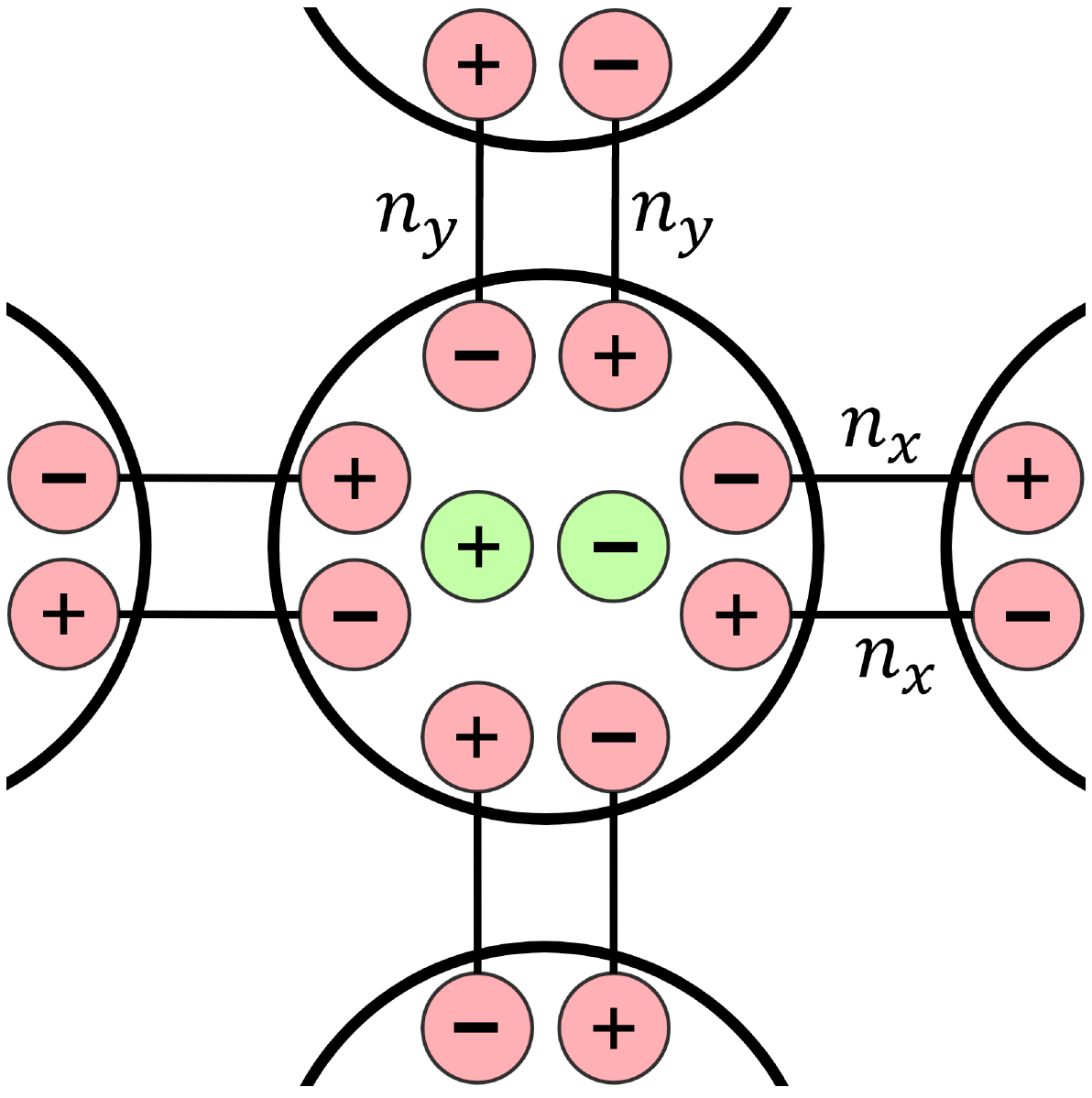}}
    \caption{Structure for the $U(1)$ symmetric TNS. Both the physical and virtual modes are categorized into two charge sectors, labeled $+$ and $-$, respectively. For $U(1)$ symmetric GfTNS, each local tensor (large black circle) again corresponds to a Gaussian state coupling physical and virtual modes, but the coupling only exists between $+$ and $-$ charge sectors, so as to ensure the desired charge conservation. The maximally entangled pair state (solid line) also only couples virtual states from distinct charge sectors. For each charge sector, there are $n_x$ ($n_y$) complex fermionic modes in the $L/R$ ($D/U$) virtual sector of each local tensor.}
    \label{fig: U(1) symmetric structure}
\end{figure}

The GfTNS construction for the $U(1)$-symmetric state proceeds in a fashion analogous to that of the BCS state discussed in Sec. \ref{sec: GfTNS review}. Similar to Eq. \eqref{eq: fiducial state without U(1)}, we begin with a tensor network state of local physical and virtual modes formulated in real space:
\begin{equation}
    \ket{\psi_{vp}} \propto e^{\frac{1}{2}\widetilde{\mathscr{N}}_{ij} a^\dagger_i a^\dagger_j} \ket{0}_v \ket{0}_p.
\end{equation}
The creation operator $a^\dagger$ includes both physical ($c^\dagger$) and virtual ($v^\dagger$) modes: $a^\dagger = \begin{pmatrix}
    c^\dagger \\ v^\dagger
\end{pmatrix}$, with the indices $i$ and $j$ carrying information about both the site position $\br$ as well as the orbital information. As before, $\ket{0}_p$ and $\ket{0}_v$ represent the vacuum for the physical and virtual modes, respectively. In anticipation of imposing the $U(1)$ charge conservation, which will become clear after we perform a particle-hole transformation on the resulting TNS, it is crucial to further introduce two charge sectors~\cite{Li2023GaussianPEPS}, denoted as $(+)$ and $(-)$, for both the physical and virtual modes, namely, $c^\dagger=\begin{pmatrix}
    c^\dagger_+ \\ c^\dagger_-
\end{pmatrix}$ and $v^\dagger=\begin{pmatrix}
    v^\dagger_+ \\ v^\dagger_-
\end{pmatrix}$. In order to construct the state in Eq. \eqref{eq: U(1)-symmetric Gaussian state in k-space}, the $n^{(\pm)}_p$ empty/occupied physical orbitals in the physical reference state $\ket{11...100...0}_p$ are assigned to the $(\pm)$-sector, respectively. As for the virtual modes, similar to Sec. \ref{sec: GfTNS review}, each of the virtual sector contains $L$, $R$, $D$, and $U$, modes so that $(v_\pm^\dagger)^T=(v_{\pm,L}^\dagger, v_{\pm,R}^\dagger, v_{\pm,D}^\dagger,v_{\pm,U}^\dagger)$, with $n_x$ modes contained in  $v_{\pm,L/R}^\dagger$, and $n_y$ modes in $v_{\pm,U/D}^\dagger$. As before, $n_{x,y}$ control the bond dimensions. Altogether, each local tensor contains $n^{(-)}_p+n^{(+)}_p$ number of physical orbitals, and $4n_x+4n_y$ virtual modes. The complete local tensor structure, with the newly introduced $(\pm)$-sector related to the conserved $U(1)$-charge, is represented by the large circle illustrated in Fig. \ref{fig: U(1) symmetric structure}.

Next, we specify the structure of the TNS matrix $\widetilde{\mathscr{N}}$, which is again diagonal in the position basis and depends on the site-independent local tensor in the orbital basis as $\widetilde{\mathscr{N}}_{(\br,\alpha),(\br',\alpha')} = \widetilde{N}_{\alpha,\alpha'} \delta_{\br,\br'}$, with
\begin{equation}
    \widetilde{N}=\begin{pmatrix}
    \widetilde{N}^{pp} & \widetilde{N}^{pv} \\ \widetilde{N}^{vp} & \widetilde{N}^{vv}
\end{pmatrix}.
\end{equation}
To impose the $U(1)$ symmetry, we demand there to be \textit{no coupling} within the $(+)$-sector or within the $(-)$-sector, individually. The generic $U(1)$-conserving form of $\widetilde{N}$ is then given by
\begin{subequations}\label{eq: block structure of N, U(1) symmetric}
\begin{align}
 \widetilde{N}^{pp} = \begin{pNiceArray}[first-row,first-col]{c|c}
  & + & - \\
 + & 0 & N^{pp} \\
  \hline
 - & -(N^{pp})^T & 0
\end{pNiceArray},\\
\widetilde{N}^{pv} = -(\widetilde{N}^{vp})^T = \begin{pNiceArray}[first-row,first-col]{c|c}
  & + & - \\
 + & 0 & N^{pv}_+ \\
  \hline
 - & N^{pv}_- & 0
\end{pNiceArray},\\
\widetilde{N}^{vv} = \begin{pNiceArray}[first-row,first-col]{c|c}
  & + & - \\
 + & 0 & N^{vv} \\
  \hline
 - & -(N^{vv})^T & 0
\end{pNiceArray},
\end{align}
\end{subequations}
where the above forms have also been simplified using the fermionic anti-symmetry. 

To obtain the physical state, we again adopt the virtual-state projection formalism as introduced in Sec. \ref{sec: GfTNS review}. To be consistent with the $U(1)$-symmetry, we adopt the following virtual state:
\begin{equation}\label{eq: M-block structure, U(1)-symmetric}
    \ket{\psi_v} \propto e^{\frac{1}{2}\widetilde{M}_{ij} v^\dagger_i v^\dagger_j}\ket{0}_v,\quad \widetilde{M} = \begin{pNiceArray}[first-row,first-col]{c|c}
  & + & - \\
 + & 0 & M \\
  \hline
 - & -M^T & 0
\end{pNiceArray},
\end{equation}
where $M$ has been introduced earlier in Eq.~\eqref{eq: real-space M} to define the entangled pair states in the virtual reference state in the BCS case. Our reference state here for the $U(1)$-symmetric construction is thus analogous to Eq.~\eqref{eq: reference virtual state}, but once again the requirement of $U(1)$ symmetry imposes a block structure on $\widetilde{M}$ that encodes the specific correlation of the entangled pair states between the $(+)$ and $(-)$ charge sectors. The correlation in the virtual reference state $\ket{\psi_v}$ is depicted by the solid lines in Fig. \ref{fig: U(1) symmetric structure}. Projecting $\ket{\psi_{vp}}$ onto $\ket{\psi_v}$ by following the recipe detailed in Appendix~\ref{appendix: virtual state}, we obtain
\begin{equation}
    \braket{\psi_v|\psi_{vp}} \propto e^{\frac{1}{2}\widetilde{M}^p_{ij} c^\dagger_i c^\dagger_j} \ket{0}_p
\end{equation}
with
\begin{equation}\label{eq: real_space_relation, U(1)-symmetric} 
    \widetilde{M}^p = \widetilde{\mathscr{N}}^{pp} + \widetilde{\mathscr{N}}^{pv}[\widetilde{\mathscr{N}}^{vv}+(\widetilde{M}^\dagger)^{-1}]^{-1} (\widetilde{\mathscr{N}}^{pv})^{T}.
\end{equation}
Importantly, with the imposed block structure (at the level of $(\pm)$-charge sector) as illustrated in Eqs.~\eqref{eq: block structure of N, U(1) symmetric} and ~\eqref{eq: M-block structure, U(1)-symmetric}, $\widetilde{M}^p$ also acquires the same block structure:
\begin{equation}\label{eq: Mp-block structure, U(1)-symmetric}
 \widetilde{M}^p = \begin{pNiceArray}[first-row,first-col]{c|c}
  & + & - \\
 + & 0 & M^p_+ \\
  \hline
 - & M^p_- & 0
\end{pNiceArray},
\end{equation}
where one can explicitly check that $M^p_- = -(M^p_+)^T$, as is consistent with the fermionic anti-symmetry. For a translation-invariant TNS construction, as we have been assuming all along in this work, the Fourier transform of $M^p_+$ takes the following form in the momentum space:
\begin{equation}
    M^p_+(\bk) = N^{pp} + N^{pv}_+ [N^{vv}+M(\bk)]^{-1}(N^{pv}_-)^T, 
\end{equation}
where $M(\bk)$ has been given in Eq.~\eqref{eq: Mk_2D}.
Other than the introduction of the $(\pm)$-charge sectors, the above expression is analogous to Eq.~\eqref{eq: k-space formula for M^p} for the GfTNS construction of BCS states without $U(1)$ symmetry, whose detailed derrivation can be found in Appendix~\ref{appendix: fourier}. Correspondingly, the momentum-space representation of the result of the virtual-state projection is
\begin{equation}\label{eq: U(1) construction before PH}
    \braket{\psi_v|\psi_{vp}} \propto \prod_\bk e^{[M^p_+(\bk)]_{\bar{\beta}\beta} c^\dagger_{(-\bk,\bar{\beta})} c^\dagger_{(\bk,\beta)}} \ket{0}_p.
\end{equation}

As the last step of our construction, in order to obtain a $U(1)$-symmetric state, we perform a particle-hole transformation on the state in Eq. ~\eqref{eq: U(1) construction before PH} such that every physical orbital in the $(-)$-charge sector (which are labeled by $\beta$) transforms as $c^\dagger_{(\bk,\beta)} \mapsto c_{(-\bk,\beta)}$, and the physical reference state transforms accordingly from the absolute vacuum $\ket{0}_p$ to $\ket{11...100...0}_p$, where the latter has been defined in Eq.~\eqref{eq: U(1)-symmetric Gaussian state in k-space}. Introducing $M^p(\bk) \equiv M^p_+(-\bk)$, our GfTNS construction thus produces the following Gaussian state:
\begin{subequations}\label{eq: U(1) construction after PH}
\begin{align}
    \ket{\psi_p} &\propto \prod_\bk e^{ [M^p(\bk)]_{\bar{\beta} \beta}c^\dagger_{(\bk,\bar{\beta})} c_{(\bk,\beta)}} \vert\underbrace{11...1}_{n^{(-)}_p}\underbrace{00...0}_{n^{(+)}_p}\rangle_p, \\
    M^p(\bk) &= N^{pp} + N^{pv}_+ [N^{vv}+M(-\bk)]^{-1}(N^{pv}_-)^T. 
\end{align}
\end{subequations}
This completes the $U(1)$-symmetric GfTNS construction for a generic 2D band insulator with $n^{(-)}_p$ filled bands and $n^{(+)}_p$ empty bands.

\subsubsection{Gauge redundancy}

Similar to our discussion of the GfTNS construction for BCS states in Sec.~\ref{sec: GfTNS review}, there are also gauge redundancies in the $U(1)$-symmetric TNS construction described above. The local tensor can be transformed by a conjugation, under which the virtual basis $v^\dagger_{(\br,\nu)}$ on each horizontal (vertical) leg is rotated by a Gaussian operator $e^{H_x}$ ($e^{H_y}$), which is parametrized by a quadratic $H_{\rho}$ ($\rho=x,y$):
\begin{equation}
\begin{split}
    H_\rho = \sum_{i,j=1}^{n_\rho} v^\dagger_{+,i}[H_\rho^{(1)}]_{ij}v^\dagger_{-,j}+v_{+,i}[H_\rho^{(2)}]_{ij}v_{-,j}\\
    +v^\dagger_{+,i} [H_\rho^{(3)}]_{ij} v_{+,j}+v^\dagger_{-,i}[H_\rho^{(4)}]_{ij}v_{-,j}.
\end{split}
\end{equation}
Recall that we have introduced the $(\pm)$-charge sector to aid our $U(1)$-symmetric TNS construction, and the above form indeed provides the most general transformation that preserves this $U(1)$ symmetry. Here $H_\rho^{(1,2,3,4)}$ are both generic $n_\rho \times n_\rho$ complex matrices, hence gauge redundancies are parametrized by $4n_x^2+4n_y^2$ complex parameters.

In the rest of this section, we focus on a generic two-band insulator with one filled band, i.e., $n^{(-)}_p = n^{(+)}_p = 1$. The total number $\mathcal{N}$ of independent TNS parameters accounting for both matrices $N^{pp}$, $N^{pv}_{\pm}$, $N^{vv}$, and upon eliminating all gauge redundancies, is given by
\begin{equation}\label{eq: TNS counting, U(1) case}
    \mathcal{N} = 8n_xn_y+4n_x+4n_y+1.
\end{equation}
To be specific, we restrict to the following form for the TNS matrix $N^{vv}$ that specifies the local virtual-mode couplings:
    \begin{equation}\label{eq: U(1) form of Nvv}
    N^{vv} = \begin{pNiceArray}[first-row,first-col]{cc|cc}
  & L & R & D & U \\
  L & \mathbf{0}_{n_x} & N^{vv}_{LR} & N^{vv}_{LD} & N^{vv}_{LU} \\
  R & N^{vv}_{RL} & \mathbf{0}_{n_x} & N^{vv}_{RD} & N^{vv}_{RU} \\
  \hline
  D & N^{vv}_{DL} & N^{vv}_{DR} & \mathbf{0}_{n_y} & N^{vv}_{DU} \\
  U & N^{vv}_{UL} & N^{vv}_{UR} & N^{vv}_{UD} & \mathbf{0}_{n_y}
\end{pNiceArray}
\end{equation}
where the sub-blocks $N^{vv}_{LR},N^{vv}_{RL},$ $N^{vv}_{DU},\; \text{and}\;N^{vv}_{UD}$ are chosen to be diagonal matrices:
\begin{equation}
    N^{vv}_{\gamma\delta} = \begin{pNiceMatrix}[nullify-dots,xdots/line-style=loosely dotted]
[N^{vv}_{\gamma\delta}]_1 & \hspace*{1cm} & 0 \\[8mm]
& \Ddots & \\
0 & & [N^{vv}_{\gamma\delta}]_{n_\rho}
\end{pNiceMatrix}
\end{equation}
for $\rho=x,y$ and $\gamma\delta\in\{LR, RL, DU, UD\}$. In this form, $N^{vv}$ contains $8n_xn_y+2n_x+2n_y$ independent parameters. Furthermore, we choose $N^{pv}_-=(\ [1]_{n_x}\ |\ [1]_{n_x}\ |\ [1]_{n_y}\ |\ [1]_{n_y}\ )$, so that $N^{pv}_+$ and $N^{pp}$ contribute the remaining $2n_x+2n_y+1$ parameters. Our main interest here is the expressivity of such $U(1)$-symmetric GfTNS states.

\subsection{Momentum-space wavefunction and trigonometric polynomials}\label{subsec: U(1) Trig Polynomials}
Focusing on $n^{(-)}_p = n^{(+)}_p = 1$ and following Eq.~\eqref{eq: U(1) construction after PH}, it is straightforward to see that the momentum-space wavefunction $M^p(\bk)$ is generally a ratio of trigonometric polynomials (TPs): $M^p(\bk) = v(\bk)/u(\bk)$, with $u(\bk),v(\bk) \in \mathbb{C}[e^{\pm ik_x}, e^{\pm i k_y}]$. However, notice that there is an important difference from the form of $M^p(\bk)$ in the BCS case as given by Eqs.~\eqref{eq: Mp as a ratio of TP} and ~\eqref{eq: general TPs}, since for the $U(1)$-symmetric case considered here there is no additional symmetry constraint on $M^p(\bk)$. Hence, for a generic two-band insulator given by Eq.~\eqref{eq: U(1)-symmetric Gaussian state in k-space}, the form of the momentum-space wavefunction should be taken generically as 
\begin{equation}\label{eq: ratio of TP, U(1) case}
    M^p(\bk) = \frac{v(\bk)}{u(\bk)},\; 
    \begin{cases}
        u(\bk) = \sum_{\bmm \in \mathcal{L}(d_x,d_y)}u_\bmm e^{i\bmm\cdot\bk} \\
        v(\bk) = \sum_{\bmm \in \mathcal{L}(d_x,d_y)}v_\bmm e^{i\bmm\cdot\bk}
    \end{cases},
\end{equation}
where $u_\bmm, v_\bmm \in \mathbb{C}$. Thus $u(\bk)$ and $v(\bk)$ are generic TPs with degree $(d_x,d_y)$ without any parity constraints. The integer lattice point $\bmm$ is defined to be summed over the following lattice 
\begin{equation}\label{eq: U(1) lattice points}
    \mathcal{L}(d_x,d_y) := \big\{\bmm=(m_x,m_y)\in \mathbb{Z}^2 \big\vert -d_{x,y} \leq m_{x,y} \leq d_{x,y} \big\}
\end{equation}
We caution the reader that this definition for the current study of $U(1)$-symmetric 2-band insulators is different from the other $\mathcal{L}(d_x,d_y)$ defined earlier in Eq.~\eqref{eq: non-trivial lattice  points} for studying BCS states. For a $U(1)$-symmetric state where its momentum-space wavefunction is generically a ratio of TPs of degree $(d_x,d_y)$, as in Eq.~\eqref{eq: ratio of TP, U(1) case}, it is clear that such a state is defined by $2\abs{\mathcal{L}(d_x,d_y)}-1 = 8d_xd_y+4d_x+4d_y+1$ number of 
complex parameters (with the $-1$ accounting for the overall scaling of both $u$ and $v$ that leaves $M^p$ unchanged). Comparing with Eq.~\eqref{eq: TNS counting, U(1) case}, we see that
\begin{equation}
    \mathcal{N} = 2\abs{\mathcal{L}(n_x,n_y)}-1.
\end{equation}
This is an exact match between the counting of TNS parameters with a total of $4n_{x}+4n_y$ virtual modes in each local tensor and the counting of parameters that define the momentum-space wavefunction $M^p(\bk)$ of a 2-band insulator in the form of a ratio of TPs with degree $(n_x,n_y)$. A natural question motivated by this counting, which is similar to the one we asked for BCS states in Sec. \ref{sec: GfTNS review} and addressed in Sec. \ref{sec: single-band BCS}, is whether for any generic 2-band insulator whose $M^p(\bk)$ is a ratio of TPs with degree $(n_x,n_y)$ the GfTNS construction presented in Sec.~\ref{sec: U(1)-symmetric TNS construction} is expressive enough to represent it. We return to addressing this question in Sec. \ref{sec: U(1) insulator, expressivity}.

\subsection{Wannier-type orbitals} \label{sec: WTO in U(1) symmetric case}
Next, we consider occupied orbitals that are localized in real space. 
For the $U(1)$-symmetric Gaussian state
\begin{equation}\label{eq: U(1)-symmetric Gaussian state, 2-band}
    \ket{\psi_p} \propto \prod_\bk (1+M^p(\bk)c^\dagger_{(\bk,2)}c_{(\bk,1)})\ket{1 0}_p.
\end{equation}
When $M^p(\bk)$ is a ratio of TPs with a finite degree, these orbitals are compactly supported in real space, with a size determined by the TP degree. Expressing $M^p(\bk)=v(\bk)/u(\bk)$, we can easily check that $\ket{\psi_p}$ is annihilated by the following operator
\begin{equation}
    \gamma^\dagger_\bk \equiv u(\bk)c^\dagger_{(\bk,1)}+v(\bk)c^\dagger_{(\bk,2)},
\end{equation}
This describes the occupied band. Relatedly, as we are focusing on two-band models, there must be an orthogonal mode, with $\eta_\bk \equiv v(\bk)c_{(\bk,1)}-u(\bk)c_{(\bk,2)}$ and $\{\gamma^\dagger_\bk, \eta_\bk\} =0$, that is the annihilation operator of the empty band, i.e., $\eta_\bk \ket{\psi_p}= 0 $.

Let us focus on the occupied states. Taking the Fourier transform of $\gamma^\dagger_\bk$, using $c^\dagger_{(\bk,\alpha)} = \frac{1}{\sqrt{N_s}}\sum_\br e^{i\bk\cdot\br}c^\dagger_{(\br,\alpha)}$ with $N_s$ the number of physical sites, we obtain the compactly supported WTO:
\begin{equation}\label{eq: WTO in the U(1) case}
\begin{split}
    \Gamma^\dagger_\bR &= \frac{1}{\sqrt{N_s}}\sum_\bk e^{-i\bk\cdot\bR}\gamma^\dagger_\bk \\
    &=\sum_\br \left[\tilde{u}(\br)c^\dagger_{(\bR+\br,1)}+\tilde{v}(\br)c^\dagger_{(\bR+\br,2)}\right]
\end{split}
\end{equation}
where $\tilde{u}(\br)$ and $\tilde{v}(\br)$ have been defined earlier in Eq.~\eqref{eq: def. of u and v in real space}. With $u(\bk)$ and $v(\bk)$ being TPs with Laurent degrees $(n_x,n_y)$, it is clear that $\Gamma^\dagger_\bR$ has a compact support of size $(2n_x+1)\times (2n_y+1)$. 

\subsubsection{Unique maximally localized WTO}
Readers would notice that our analysis for the WTO here in the $U(1)$-symmetric case is closely analogous to that for the BCS case in Sec.~\ref{subsec: WTO, BCS}. There are only two differences. For the $U(1)$-symmetric states, (1) the WTOs do not mix between particle and hole states (see Eq.~\eqref{eq: WTO in the U(1) case} versus Eq.~\eqref{eq: def. of WTO}), and (2) the TPs $u(\bk)$ and $v(\bk)$ (and hence their Fourier transforms) do not have any parity constraints (see Eq.~\eqref{eq: ratio of TP, U(1) case} versus Eq.~\eqref{eq: general TPs}). 

However, none of these aforementioned differences are relevant for understanding the size of WTOs. Particualrly, one can clearly see that the discussion on the maximally localized WTO at the end of Sec.~\ref{subsec: WTO, BCS} applies equally well to our 2-band $U(1)$-symmetric case, because no assumptions about symmetry properties of $u(\bk)$ and $v(\bk)$ have been invoked there. The fact that the regular polynomial ring $\mathbb{C}[e^{ik_x}, e^{ik_y}]$ is a unique factorization domain (UFD) implies that there must exist a unique maximally localized WTO (up to translation), which is associated to the coprime pair of regular TPs whose ratio equals to $M^p(\bk)$. We refer interested readers to the end of Sec.~\ref{subsec: WTO, BCS} for the detailed argument. 

Denoting the size of this maximally localized WTO as $(\mathcal{M}_x+1)\times(\mathcal{M}_y+1)$, the property of UFD implies that any WTO that annihilates the $U(1)$-symmetric Gaussian state $\ket{\psi_p}$ in Eq.~\eqref{eq: U(1)-symmetric Gaussian state, 2-band}, with a support of size $\mathcal{W}_x\times\mathcal{W}_y$, must obey $\mathcal{W}_{x,y}\geq \mathcal{M}_{x,y}+1$.

\section{Expressivity with $U(1)$ symmetry}\label{sec: U(1) insulator, expressivity}

We now move on to discuss the expressivity of GfTNS for a generic two-band insulator with one occupied band. Following the notation introduced in the previous section, such states can be represented as
\begin{equation}\label{eq: U(1)-symmetric Gaussian state, 2-band, exponential form}
    \ket{\psi_p} \propto \prod_\bk e^{M^p(\bk)c^\dagger_{(\bk,2)}c_{(\bk,1)}}\ket{1 0}_p
\end{equation}
where $c^\dagger_{(\bk,\sigma)}$ creates a fermion in orbital $\sigma(=1,2)$ at momentum $\bk$, and $\ket{10}_p$ represents a reference physical many-body state where orbital $\sigma=1$ is occupied for all $\bk$ and orbital $\sigma=2$ is empty for all $\bk$. Unlike the BCS case considered before, $M^p(\bk)$ (which we refer to as the momentum-space wavefunction) does not obey any symmetry constraint. In this section, we establish the equivalence between the following four statements regarding the $U(1)$-symmetric Gaussian fermionic state in Eq.~\eqref{eq: U(1)-symmetric Gaussian state, 2-band, exponential form}:

\begin{enumerate}[label=(\textbf{P\arabic*}'):, ref=(P\arabic*')]
    \item \textit{Area law of R{\'e}nyi-0 entropy:} For any region $A$ of size $l_x\times l_y$ unit cells, the R{\'e}nyi-0 entanglement entropy satisfies $S_0(A)\le 4n_xl_y + 4n_yl_x$. \label{itm: P1'}
    \item \textit{Compactly supported Wannier-type orbital:} There is a compactly supported Wannier-type orbital of size $(2n_x+1)\times (2n_y+1)$, which is occupied in $\ket{\psi_p}$. \label{itm: P2'}
    
    \item \textit{Momentum-space wavefunction:} $M^p(\bk)$ can be represented as a ratio of  trigonometric polynomials $u(\bk)$ and $v(\bk)$ of Laurent degree $(n_x,n_y)$, where $u(\bk), v(\bk)$ and $M^p(\bk)$ have no symmetry constraints. \label{itm: P3'}
    \item \textit{Tensor network representation:} The state is represented by a GfTNS with $2n_x$ and $2n_y$ virtual complex fermions on each horizontal and vertical leg of the tensor, respectively; correspondingly, the horizontal (vertical) bond dimension is $2^{2n_x}$ ($2^{2n_y}$). The detailed structure of the $U(1)$-symmetric TNS has been introduced in Sec. \ref{sec: U(1) TNS review}. \label{itm: P4'}
\end{enumerate}
Similar to the previous BCS discussion in Sec.~\ref{sec: single-band BCS}, the equivalence among the above four properties can be proved by showing \ref{itm: P4'} $\implies$ \ref{itm: P1'} $\implies$ \ref{itm: P2'} $\implies$ \ref{itm: P3'} $\implies$ \ref{itm: P4'}. Notably, our discussion in Sec.~\ref{subsec: EE, BCS} is equally applicable to $U(1)$-symmetric GfTNS, since the virtual-state projection formalism is identical to the case without $U(1)$. The only difference from having a $U(1)$ symmetry is the need to introduce $(\pm)$-charge sectors. We thus easily obtain \ref{itm: P4'} $\implies$ \ref{itm: P1'} from our previous discussion. We now move on to the remaining relations. 

\subsection{\ref{itm: P1'} $\implies$ \ref{itm: P2'} $\implies$ \ref{itm: P3'} }
The argument here is very similar to the BCS case. We shall again refer to Fig.~\ref{fig: GfTNS structure}(b) and proceed with only small adaptations. Consider a region $A$ of size $(2n_x+1)\times l_y$, which contains $n_p(2n_x+1)l_y = 2(2n_x+1)l_y=N_A$ number of physical fermionic modes. Assuming \ref{itm: P1'}, we have 
\begin{equation}
    S_0(A) \leq 4n_x l_y+4n_y l_x \leq  N_A + 4n_y(2n_x+1) - 2l_y
\end{equation}
Hence for $l_y>2n_y(2n_x+1)$, we have $S_0(A) < N_A$. Combining this with the Gaussian property of the state (i.e., its entanglement Hamiltonian must be quadratic and also $U(1)$-symmetric), this implies the existence of a single-particle state, compactly supported entirely within region $A$, which is either completely occupied or empty in $\ket{\psi_p}$ and thus does not contribute to the entanglement. Specifically for two-band models, as discussed in Sec.~\ref{sec: WTO in U(1) symmetric case}, this further implies the existence of an occupied WTO that is compactly supported entirely within region $A$. This WTO has a size $(2n_x+1)\times l_y$. Following our discussion about maximally localized WTO in Sec.~\ref{sec: WTO in U(1) symmetric case}, we have $2n_x+1 \geq \mathcal{M}_x+1$, where $\mathcal{M}_a+1$ is the size of the maximally localized Wannier-type orbital along the $a$-direction. Repeating the above analysis by considering a region of size $l_x \times (2n_y+1)$ with $l_x > 2n_x(2n_y+1)$, we conclude that $2n_y+1 \geq \mathcal{M}_y+1$. Hence, altogether, we conclude there exists a compactly supported WTO of size $(2n_x+1)\times(2n_y+1)$ that is occupied in $\ket{\psi_p}$, i.e. \ref{itm: P1'} $\implies$ \ref{itm: P2'}.

Denoting the creation operator for this compactly supported WTO as $\Gamma^\dagger_\bR$, its Fourier transform into the momentum space would take the form the form $u(\bk)c^\dagger_{(\bk,1)}+ v(\bk)c^\dagger_{(\bk,2)}$ where $u(\bk)$ and $v(\bk)$ are TPs with Laurent degrees $(n_x,n_y)$. From $\Gamma^\dagger_\bR \ket{\psi_p}=0$ for all $\bR$, this implies
\begin{equation}
\begin{split}
        &(u(\bk)c^\dagger_{(\bk,1)}+ v(\bk)c^\dagger_{(\bk,2)})(1+M^p(\bk)c^\dagger_{(\bk,2)}c_{(\bk,1)})\ket{10}_p=0 \\
        &\implies M^p(\bk) = \frac{v(\bk)}{u(\bk)},
\end{split}
\end{equation}
for all $\bk$. Thus, \ref{itm: P2'} $\implies$ \ref{itm: P3'}.

\subsection{\ref{itm: P3'} $\implies$ \ref{itm: P4'}} Now we return to the question raised in Sec.~\ref{subsec: U(1) Trig Polynomials} of whether the GfTNS construction is expressive enough to represent $M^p(\bk)$. We proceed similarly to the method shown for BCS states in Sec.~\ref{sec: P3 implies P4} by showing number of independent parameters required to specify $M^p(\bk)$, $\mathcal{N} = 8n_xn_y+4n_x+4n_y+1$, is equal to the number of independent parameters of a 2D GfTNS with $4n_x+4n_y$ virtual modes in each local tensor. Then, recalling from Sec.~\ref{sec: U(1)-symmetric TNS construction} the relation between the $U(1)$-symmetric momentum-space wavefunction and the TNS parameters
\begin{equation}\label{eq: U(1) relation between the momentum-space wavefunction and the tensor network parameters}
    M^p(\bk) = N^{pp} + N^{pv}_+ [N^{vv}+M(-\bk)]^{-1}(N^{pv}_-)^T,
\end{equation}
we again seek to show that Eq.~\eqref{eq: U(1) relation between the momentum-space wavefunction and the tensor network parameters} can represent any generic ratio of TPs in the form of \ref{itm: P3'}.

In comparison with the BCS case, the matrices $N^{pp},\; N^{pv}_+,\; N^{pv}_-,$ and $N^{vv}$ in Eq. ~\eqref{eq: U(1) relation between the momentum-space wavefunction and the tensor network parameters} no longer obey any symmetry constraints. This is because these matrices only represent the plus-minus sector of $\widetilde{N}$ and the fermion anti-commutation instead results in anti-symmetry constraints on the minus-plus sector. Therefore, after eliminating all gauge redundancies, the structure of $N^{vv}$, which is repeated below for the reader's convenience, 
\begin{equation}\label{eq: Copy of U(1) form of Nvv}
    N^{vv} = \begin{pNiceArray}[first-row,first-col]{cc|cc}
  & L & R & D & U \\
  L & \mathbf{0}_{n_x} & N^{vv}_{LR} & N^{vv}_{LD} & N^{vv}_{LU} \\
  R & N^{vv}_{RL} & \mathbf{0}_{n_x} & N^{vv}_{RD} & N^{vv}_{RU} \\
  \hline
  D & N^{vv}_{DL} & N^{vv}_{DR} & \mathbf{0}_{n_y} & N^{vv}_{DU} \\
  U & N^{vv}_{UL} & N^{vv}_{UR} & N^{vv}_{UD} & \mathbf{0}_{n_y}
\end{pNiceArray},
\end{equation}
is such that $N^{vv}_{LR},\; N^{vv}_{RL},\; N^{vv}_{DU},$ and $N^{vv}_{UD}$ are all independent diagonal matrices contributing a total of $2n_x+2n_y$ parameters while all of $\{N^{vv}_{LD}, N^{vv}_{DL}, N^{vv}_{LU}, N^{vv}_{UL}, N^{vv}_{RD}, N^{vv}_{DR}, N^{vv}_{RU}, N^{vv}_{UR}\}$ are generic matrices contributing $n_xn_y$ independent parameters each. 

Having also used the gauge redundancy to choose all elements of $N^{pv}_-$ to be $1$, the remaining matrices $N^{pp}$ and $N^{pv}_+$ each contribute $1$ and $2n_x+2n_y$ independent parameters respectively. Adding these together, we have the total $8n_xn_y+4n_x+4n_y+1$ independent parameters of the TNS to specify $M^p(\bk)$. We note that this results in about twice as many independent parameters as the BCS case, due to the incorporation of $\pm$ charge sectors. Relatedly, we note that this matches the counting of independent TP coefficients that defines the momentum space wavefunction $M^p(\bk)=v(\bk)/u(\bk)$, because there is no parity constraint on the TPs $u(\bk)$ and $v(\bk)$, see the discussion around Eq.~\eqref{eq: ratio of TP, U(1) case}. Other than this quantitative difference, the structure of TNS representation is qualitatively similar to our previous BCS discussion. Thus, we take the same approach to show that the TP coefficients can be independently tuned in a local region of the TNS parameter space and utilize Chevally's Theorem to extend to the full parameter space.

The specific region of TNS parameter space we focus on is again where all the $n_x\times n_y$ matrices that appear in the upper right and lower left blocks of $N^{vv}$ in Eq.~\eqref{eq: Copy of U(1) form of Nvv} are close to zero. By Taylor expansion, we derive expressions for $u(\bk)$ and $v(\bk)$ to quadratic order in $\{N^{vv}_{LD}, N^{vv}_{DL}, N^{vv}_{LU}, N^{vv}_{UL}, N^{vv}_{RD}, N^{vv}_{DR}, N^{vv}_{RU}, N^{vv}_{UR}\}$. We note again that now we must separately distinguish between $N^{vv}_{\gamma\delta}$ and $N^{vv}_{\delta\gamma}$ since they are no longer related by anti-symmetry as in the BCS case. The resulting expressions for $u(\bk)$ and $v(\bk)$ when expanding to quadratic order are shown below. The complete derivation follows the same steps described in Appendix ~\ref{subsec: Lagrange w/o U(1)}.

\begin{widetext}
\begin{table*}[t]
\caption{List of $U(1)$-symmetric 2D tensor network state (TNS) parameters used to tune the trigonometric polynomials (TPs), $u(\bk)$ and $v(\bk)$, based on Eq.~\eqref{main text eq: full u(k) and v(k) for U(1)}. Gauge redundancies of the TNS have again been used to restrict the form of $N^{vv}$ as described in Sec.~\ref{sec: U(1)-symmetric TNS construction}, and the indices $\lambda$ and $\mu$ run from $1$ to $n_x$ and $1$ to $n_y$ respectively as before.}\label{table: TNS parameters with U(1)}
\renewcommand{\arraystretch}{2} 

\begin{ruledtabular}

\begin{tabular}{c|c|c}
\text{\small TP}& $u(\bk)$ & $v(\bk)$ \\ \hline
 & $[N^{vv}_{LR}]_{\lambda}, \; [N^{vv}_{RL}]_{\lambda}, \; [N^{vv}_{DU}]_{\mu}, \; [N^{vv}_{UD}]_{\mu}, \; $  & $[N^{pv}_{+L}]_\lambda,\;[N^{pv}_{+R}]_\lambda,\;[N^{pv}_{+D}]_\mu,\;[N^{pv}_{+U}]_\lambda,\;$ \\
\text{\small TNS} & $[N^{vv}_{LU}]_{\lambda \mu}[N^{vv}_{DR}]_{\mu \lambda}, \;[N^{vv}_{RD}]_{\lambda \mu}[N^{vv}_{UL}]_{\mu \lambda}$  & $[N^{pv}_{+D}]_\mu[N^{vv}_{UR}]_{\mu\lambda}+[N^{pv}_{+R}]_\lambda [N^{vv}_{LD}]_{\lambda\mu},\;[N^{pv}_{+U}]_\mu[N^{vv}_{DR}]_{\mu\lambda}+[N^{pv}_{+R}]_\lambda [N^{vv}_{LU}]_{\lambda\mu},$ \\
&$[N^{vv}_{LD}]_{\lambda \mu}[N^{vv}_{UR}]_{\mu \lambda}, \; [N^{vv}_{RU}]_{\lambda \mu}[N^{vv}_{DL}]_{\mu \lambda}$& $[N^{pv}_{+D}]_\mu[N^{vv}_{UL}]_{\mu\lambda}+[N^{pv}_{+L}]_\lambda [N^{vv}_{RD}]_{\lambda\mu},\;[N^{pv}_{+U}]_\mu[N^{vv}_{DL}]_{\mu\lambda}+[N^{pv}_{+L}]_\lambda [N^{vv}_{RU}]_{\lambda\mu},$\\
\end{tabular}

\end{ruledtabular}
\end{table*}

\begin{equation}\label{main text eq: full u(k) and v(k) for U(1)}
    \begin{split}
        &u(\bk) = (-1)^{n_x+n_y}\prod_{\lambda=1}^{n_x}\prod_{\mu=1}^{n_y}\; (e^{-ik_x}-[N^{vv}_{LR}]_\lambda)(e^{+ik_x}+[N^{vv}_{RL}]_\lambda)(e^{-ik_y}-[N^{vv}_{DU}]_\mu)(e^{+ik_y}+[N^{vv}_{UD}]_\mu)\\
    &\quad\quad\quad\quad\times \left(1+\sum_{\lambda=1}^{n_x}\sum_{\mu=1}^{n_y}\left[\frac{[N^{vv}_{RD}]_{\lambda\mu}[N^{vv}_{UL}]_{\mu\lambda}}{(e^{+ik_y}+[N^{vv}_{UD}]_\mu)(e^{+ik_x}+[N^{vv}_{RL}]_\lambda)}+\frac{(-1)^{n_x}[N^{vv}_{LD}]_{\lambda\mu}[N^{vv}_{UR}]_{\mu\lambda}}{(e^{+ik_y}+[N^{vv}_{UD}]_\mu)(e^{-ik_x}-[N^{vv}_{LR}]_\lambda)}\right.\right.\\
        &\hspace{3cm}\left.\left. +\frac{(-1)^{n_y}[N^{vv}_{RU}]_{\lambda\mu}[N^{vv}_{DL}]_{\mu\lambda}}{(e^{-ik_y}-[N^{vv}_{DU}]_\mu)(e^{+ik_x}+[N^{vv}_{RL}]_\lambda)}+\frac{(-1)^{n_x+n_y}[N^{vv}_{LU}]_{\lambda\mu}[N^{vv}_{DR}]_{\mu\lambda}}{(e^{-ik_y}-[N^{vv}_{DU}]_\mu)(e^{-ik_x}-[N^{vv}_{LR}]_\lambda)}\right]\right)\\
        &v(\bk)  = (-1)^{n_x+n_y}\prod_{\lambda=1}^{n_x}\prod_{\mu=1}^{n_y}\; (e^{-ik_x}-[N^{vv}_{LR}]_\lambda)(e^{+ik_x}+[N^{vv}_{RL}]_\lambda)(e^{-ik_y}-[N^{vv}_{DU}]_\mu)(e^{+ik_y}+[N^{vv}_{UD}]_\mu)\\
        &\hspace{1cm}\times\left\{N^{pp}+\sum_{\lambda=1}^{n_x}\left[\frac{[N^{pv}_{+R}]_\lambda}{([N^{vv}_{LR}]_\lambda-e^{-ik_x})}+\frac{[N^{pv}_{+L}]_\lambda}{([N^{vv}_{RL}]_\lambda+e^{+ik_x})}\right] + \sum_{\mu=1}^{n_y}\left[\frac{[N^{pv}_{+U}]_\mu}{([N^{vv}_{DU}]_\mu-e^{-ik_y})} + \frac{[N^{pv}_{+D}]_\mu}{([N^{vv}_{UD}]_\mu+e^{+ik_y})}\right] \right.\\
        &\hspace{2cm}\left.- \sum_{\lambda=1}^{n_x}\sum_{\mu=1}^{n_y}\left[ \frac{[N^{pv}_{+D}]_\mu[N^{vv}_{UR}]_{\mu\lambda}+[N^{pv}_{+R}]_\lambda[N^{vv}_{LD}]_{\lambda\mu}}{([N^{vv}_{UD}]_\mu + e^{+ik_y})([N^{vv}_{LR}]_\lambda - e^{-ik_x})} + \frac{[N^{pv}_{+U}]_\mu[N^{vv}_{DR}]_{\mu\lambda}+[N^{pv}_{+R}]_\lambda[N^{vv}_{LU}]_{\lambda\mu}}{([N^{vv}_{DU}]_\mu - e^{-ik_y})([N^{vv}_{LR}]_\lambda - e^{-ik_x})} \right.\right. \\
        &\hspace{4cm}\left.\left. + \frac{[N^{pv}_{+D}]_\mu[N^{vv}_{UL}]_{\mu\lambda}+[N^{pv}_{+L}]_\lambda[N^{vv}_{RD}]_{\lambda\mu}}{([N^{vv}_{UD}]_\mu + e^{+ik_y})([N^{vv}_{RL}]_\lambda + e^{+ik_x})}+\frac{[N^{pv}_{+U}]_\mu[N^{vv}_{DL}]_{\mu\lambda}+[N^{pv}_{+L}]_\lambda[N^{vv}_{RU}]_{\lambda\mu}}{([N^{vv}_{DU}]_\mu - e^{-ik_y})([N^{vv}_{RL}]_\lambda + e^{+ik_x})} \right]\right\}
    \end{split}
\end{equation}
\end{widetext}
Since the structure of $u(\bk)$ and $v(\bk)$ remains the same as before, we can apply an identical method of Lagrange interpolation to tune each of the TP coefficients by simply doubling the number of Lagrange points to account for the extra TP coefficients as seen in ~\eqref{eq: choices of ki for Lagrange U(1)}. The exact steps are described in Appendix~\ref{subsec: Lagrange w/ U(1)}, and a summary of the TNS parameters used to tune each TP coefficients is given in Table~\ref{table: TNS parameters with U(1)}. This shows that the independent TNS parameters can be used to independently tune each and every TP coefficient in a local region of parameter space, and by using Chevally's Theorem, the TNS can represent a generic \footnote{For the specific meaning of ``generic'', we refer readers to the remark detailed in Sec.~\ref{sec: Concluding remarks for P3 implies P4}.} half-filled two-band insulator (i.e.,  $\ket{\psi_p}$ in Eq.~\eqref{eq: U(1)-symmetric Gaussian state, 2-band, exponential form}) throughout the parameter space. This proves \ref{itm: P3'} $\implies$ \ref{itm: P4'}. 

Combined with the discussion in last section, we have arrived at the equivalence between \ref{itm: P1'}, \ref{itm: P2'}, \ref{itm: P3'} and \ref{itm: P4'}, for a generic $U(1)$ symmetric two-band Gaussian state in the form of Eq.~\eqref{eq: U(1)-symmetric Gaussian state, 2-band, exponential form}. In particular, this  establishes the expressivity of finite bond-dimension GfTNS for $U(1)$ symmetric two-band insulators in connection with its entanglement area-law bound (by \ref{itm: P1'} $\iff$ \ref{itm: P4'}).

\section{Discussion}

We have analytically characterized the expressivity of Gaussian fermionic tensor network state (GfTNS) representation for two families of 2D translation-invariant Gaussian fermionic states: single-band BCS superconductors and half-filled two-band insulators. For each family, we prove that an area law for the R{\'e}nyi-0 entropy is equivalent to having an exact Gaussian fermionic TNS representation with a finite bond dimension determined by the R{\'e}nyi-0 entropy. These results provide useful guidance for determining the minimal resources needed for simulating 2D Gaussian fermionic states, which are relevant for studying itinerant electronic systems in quantum materials.

For BCS states, see Eq.~\eqref{eq: BCS Gaussian form}, we show that when R{\'e}nyi-0 entropy satisfies $S_0(A) \leq 2n_xl_y+2n_y l_x$ ($n_{x,y} \in \mathbb{Z}^+_0$), for every rectangular region $A$ of size $l_x\times l_y$ sites, there is an exact GfTNS representation with horizontal (vertical) bond dimension $2^{n_x}$ ($2^{n_y}$). For $U(1)$-symmetric half-filled two-band Gaussian states, see Eq.~\eqref{eq: U(1)-symmetric Gaussian state, 2-band, exponential form}, we show that an area-law bound $S_0(A) \leq 4n_xl_y+4n_y l_x$ guarantees an exact $U(1)$-symmetric GfTNS representation with horizontal (vertical) bond dimension $2^{2n_x}$ ($2^{2n_y}$). In both cases, we have further connected the minimal bond dimension to the size of the maximally localized Wannier-type orbital occupied by this state: if the maximally localized Wannier-type orbital is compactly supported within $(2\mathcal{M}_x+1)\times (2\mathcal{M}_y+1)$ sites, then the corresponding minimal GfTNS is given by $(n_x,n_y) = (\mathcal{M}_x,\mathcal{M}_y)$.

Before we conclude, let us discuss possible directions to extend our current results. This work has focused on GfTNS with $n_p\leq2$ physical fermionic modes per tensor. Simulating generic multi-band systems would also require consideration for $n_p> 2$. However, for $n_p> 2$, the minimal bond dimension is no longer determined only by the R{\'e}nyi-0 entropy in contrast to what we have established for $n_p \leq 2$. , To illustrate this, we use a simple counting argument, similar to those presented in Sec.~\ref{sec: GfTNS review} and Sec.~\ref{sec: U(1) TNS review}. For concreteness, consider a $U(1)$-symmetric $n_p$-band Gaussian state with one occupied band (c.f. Eq.~\eqref{eq: U(1) construction after PH}):
\begin{equation}\label{eq: main, U(1)-symmetric Gaussian state, generic n_p}
    \ket{\psi_p} \propto \prod_\bk (1+\sum_{\sigma=2}^{n_p}M^p_\sigma(\bk)c^\dagger_{(\bk,\sigma)}c_{(\bk,1)})\ket{10...0}_p,
\end{equation}
where the physical reference state $\ket{10...0}_p$ corresponds to filling the $\sigma=1$ orbital and leaving all other orbitals $(\sigma=2,...,n_p)$ empty for all $\bk$. The GfTNS representation of such states, as discussed in Sec.~\ref{sec: U(1) TNS review}, requires $M^p_\sigma(\bk)=v_\sigma(\bk)/u(\bk)$ to be a ratio of trignometric polynomials (TPs). Denoting the polynomial degree for all involved TPs as $(\mathcal{M}_x, \mathcal{M}_y)$, we see that $\ket{\psi_p}$ is specified by (details provided in Appendix~\ref{sec: appendix, counting for np > 2})
\begin{equation}
    \mathcal{N}_p =4\mathcal{M}_x\mathcal{M}_yn_p+2(\mathcal{M}_x+\mathcal{M}_y)n_p+n_p-1
\end{equation}
number of complex parameters. On the other hand, the $U(1)$-symmetric GfTNS with horizontal (vertical) bond dimension $2^{2n_x}$ ($2^{2n_y}$) contains 
\begin{equation}
    \mathcal{N}_{\rm TNS} = 8n_xn_y+2(n_x+n_y)n_p+n_p-1
\end{equation}
number of independent TNS parameters. To ensure the GfTNS is expressive enough, it is required that $\mathcal{N}_{\rm TNS} \geq \mathcal{N}_p$, which cannot be satisfied by naively setting $n_{x,y}=\mathcal{M}_{x,y}$ for $n_p>2$. This expressivity condition is satisfied instead with 
\begin{equation}\label{eq: bond-dimension bound for general n_p}
    n_{x,y} \geq \mathcal{M}_{x,y}\sqrt{\frac{n_p}{2}}.
\end{equation}

Recall that the TP degree $\mathcal{M}_{x,y}$ has a very intuitive physical meaning: a compactly supported Wannier-type orbital occupied by $\ket{\psi_p}$ has size $(2\mathcal{M}_x+1)\times (2\mathcal{M}_y+1)$, and hence for a partition cut along the vertical (horizontal) direction, the R{\'e}nyi-0 entropy is upper bounded by $2\mathcal{M}_x l_y$ ($2\mathcal{M}_y l_x$), as detailed in App.~\ref{sec: appendix, counting for np > 2}. Our counting argument, leading to Eq.~\eqref{eq: bond-dimension bound for general n_p}, indicates that for generic multi-band systems with $n_p>2$, the minimal bond dimension required by an exact GfTNS representation is \textit{not} solely given by the R{\'e}nyi-0 entropy. Despite the fact that a GfTNS with bond dimension $2^{2n_{x,y}}$ does obey an area-law bound given by $2n_{x}l_{y}$ ($2n_{y}l_{x}$) for each horizontal (vertical) partition, one should not expect the converse where every multi-band Gaussian state satisfying this area-law bound would admit an exact GfTNS representation with bond dimension $2^{2n_{x,y}}$. 
We also note that Eq.~\eqref{eq: bond-dimension bound for general n_p} is a necessary condition for expressivity, but we have not established that it is sufficient for $n_p>2$. It is interesting to investigate whether a different TNS architecture, for instance by involving multiple local tensors per unit cell (instead of just a single tensor for all $n_p$ physical orbitals), could provide a more economical representation of multi-band systems.

Lastly, let us comment on how our work is relevant for gapless fermionic systems. While we have frequently referred to superconductors and band insulators in this work, it should be emphasized that all of our results have been established with respect to the Gaussian ``states" and their entanglement properties, without reference to any Hamiltonians or associated spectral gaps. Our results thus also apply to critical gapless states that obey the entanglement area law, and become useful when the associated wavefunction $M^p(\bk)$ is a ratio of TPs of finite degrees. As a specific example, consider the two-band Bloch Hamiltonian $h_\bk =(\sin k_x-\sin k_y)\sigma_x -\cos k_x\cos k_y\,\sigma_z$. The associated momentum space wavefunction (c.f. Eq.~\eqref{eq: U(1)-symmetric Gaussian state, 2-band, exponential form}) takes the form $M^p(\bk) = (\sin k_y-\sin k_x)/(1-\sin k_x \sin k_y+\cos k_x \cos k_y)$, which is a ratio of TPs of degree $(1,1)$.  Note that the energy spectrum has a quadratic band-touching at $\pm\bk_0=\pm(\pi/2, \pi/2)$, thus the corresponding Gaussian state $\ket{\psi_p}$ describes a critical semimetal. Applying our results in this $U(1)$-symmetric context, the minimal bond dimension of its $U(1)$-symmetric GfTNS is determined to be $4$ for both the horizontal and vertical directions. Similar critical states and their TNS representations, in the BCS context, were studied in the early literature \cite{PhysRevA.81.052338}.
On the other hand, a Dirac fermion has $M^p(\bk) = (k_x + ik_y)/|\bk|$, which cannot be expressed as a ratio of two trigonometric polynomials (fixing $k_y=0$, $M^p(\bk)$ jumps at $k_x=0$); therefore, any GfTNS with finite bond dimension cannot represent a Dirac fermion in the thermodynamic limit.

Beyond critical states, which still obey an entanglement area law, an interesting but challenging extension for the study of GfTNS expressivity is to target metallic states with Fermi surfaces. These gapless states are well known to violate the entanglement area law \cite{Wolf2006, GioevKlich2006}, featuring additional logarithmic enhancement in the entanglement entropy, and hence do not admit exact TNS representation with fixed finite bond dimensions. Nevertheless, a metallic Fermi surface state can be approximated by a weakly paired BCS superconductor with a vanishing pairing gap \cite{TN_Fermi_Surface, Finite_Ent_Scaling_2023}. 
Our results relating the analytic structure of $M^p(\bk)$ (via the degree of trigonometric polynomials) to the  bond dimension allow one to quantify how sharply the Fermi surface can be resolved. This is the subject of an accompanying work \cite{TNS_metal_Dai}.

\begin{acknowledgments}
We thank Michael P. Zaletel for helpful discussions. ZD is supported by startup funds at the University of Pittsburgh. PMT is supported by postdoctoral research fellowships at the Princeton Center for Theoretical Science and the Leinweber Institute for Theoretical Physics at the University of Chicago. 
\end{acknowledgments}

\appendix

\section{Virtual-state projection}\label{appendix: virtual state}
Here we provide a detailed derivation of the real-space relation in  Eq.~\eqref{eq: real_space_relation} connecting the TNS parameters to the matrix $M^p$ that specifies the physical Gaussian state. As defined in Eq.~\eqref{eq: physical state projection}, the physical state is written in terms of the TNS and a reference virtual state as
\begin{equation}
\begin{split}
    \ket{\psi_p}=\braket{\psi_v|\psi_{vp}}&\propto\bra{0_v}e^{\frac{1}{2}M^\dag_{ij}v_iv_j}e^{\frac{1}{2}\mathscr{N}_{ij}a^\dag_ia^\dag_j}\ket{0_v}\ket{0_p}\\
    &\propto e^{\frac{1}{2}M^p_{ij}b^\dagger_i b^\dagger_j} \ket{0}_p
\end{split}
\end{equation}
The right hand side is evaluated to derive the relation of $M^p$ to the $M$ and $\mathscr{N}$ matrices. This is done by associating the creation/annihilation operators with Grassman variables, $a^\dag_i=\begin{pmatrix} b^\dag_i \\ v^\dag_i \end{pmatrix} \rightarrow \begin{pmatrix} \eta \\ \theta \end{pmatrix}$ and $v_i\rightarrow\Tilde{\theta}$ which allows the trace over the virtual modes to be evaluated as a Grassman integral over the variables $\theta$ and $\tilde{\theta}$ which correspond to the virtual creation and annihilation operators. The resulting integral form of the trace is expressed as
\begin{equation}
    \begin{split}
        &\bra{0_v}e^{\frac{1}{2}M^\dag_{ij}v_iv_j}e^{\frac{1}{2}\mathscr{N}_{ij}a^\dag_ia^\dag_j}\ket{0_v} \\
        & \propto \int D\theta D\Tilde{\theta}\ e^{\frac{1}{2}\begin{pmatrix}\eta^T&\theta^T\end{pmatrix}\begin{pmatrix}\mathscr{N}^{pp}&\mathscr{N}^{pv}\\\mathscr{N}^{vp}&\mathscr{N}^{vv}\end{pmatrix}\begin{pmatrix}\eta\\\theta\end{pmatrix}}e^{\frac{1}{2}\Tilde{\theta}^T M^\dag\Tilde{\theta}}e^{\Tilde{\theta}^T\theta}\\
        & \propto e^{\frac{1}{2}\eta^T \mathscr{N}^{pp}\eta}\int D\theta\ e^{\eta^T \mathscr{N}^{pv}\theta+\frac{1}{2}\theta^T(\mathscr{N}^{vv}+(M^\dag)^{-1})\theta}
    \end{split}
\end{equation}
where in the above we have given a block structure to the $\mathscr{N}$ matrix as done in the main text.  
Performing a change of variables $\theta_n=\mathscr{N}^{pv}\theta\rightarrow D\theta_n=(\det \mathscr{N}^{pv})^{-1}D\theta$ allows the remaining integral to be evaluated, giving the final result.
\begin{equation}
    \begin{split}
        &e^{\frac{1}{2}\eta^T M^p\eta}\propto e^{\frac{1}{2}\eta^T \mathscr{N}^{pp}\eta}\\
        &\times\int D\theta_n\ e^{\eta^T\theta_n}e^{\frac{1}{2}\theta_n^T(((\mathscr{N}^{pv})^T)^{-1}(\mathscr{N}^{vv}+(M^\dag)^{-1})(\mathscr{N}^{pv})^{-1})\theta_n}\\
        & \propto e^{\frac{1}{2}\eta^T(\mathscr{N}^{pp}+\mathscr{N}^{pv}(\mathscr{N}^{vv}+(M^\dag)^{-1})^{-1}(\mathscr{N}^{pv})^T)\eta}\\
    \end{split}
\end{equation}
Thus by comparison of the left and right sides, we conclude the resulting relation in real space
\begin{equation}
     M^p = \mathscr{N}^{pp}+\mathscr{N}^{pv}[\mathscr{N}^{vv}+(M^\dagger)^{-1}]^{-1}(\mathscr{N}^{pv})^T.
\end{equation}
This is Eq.~\eqref{eq: real_space_relation} in the maint text.

\section{Derivation of Eq.~\eqref{eq: k-space formula for M^p}}\label{appendix: fourier}
Here we provide a detailed derivation of the momentum-space relation in Eq.~\eqref{eq: k-space formula for M^p} from the real-space relation in Eq.~\eqref{eq: real_space_relation}. We  consider the Fourier transform from real to momentum space:
\begin{equation}
    a^\dagger_{(\br, \alpha)} =F_{(\br, \alpha), (\bk, \alpha')}a^\dagger_{(\bk, \alpha')},
\end{equation}
where $F_{(\br, \alpha), (\bk, \alpha')} = \frac{1}{\sqrt{N_{s}}}e^{-i\br\cdot \bk} \delta_{\alpha,\alpha'}$, and $N_s$ is the total number of sites. Note that $F$ is a unitary transformation that is diagonal in the orbital space, and also satisfies
\begin{equation}
\begin{split}
    &[F F^T]_{(\br, \alpha), (\br', \alpha')} = \delta_{\br, -\br'} \delta_{\alpha,\alpha'},\\
    &[F^T F]_{(\bk, \alpha), (\bk', \alpha')} = \delta_{\bk, -\bk'} \delta_{\alpha,\alpha'}.
\end{split}
\end{equation}
The Fourier transform of $M^p$ is constructed as
\begin{align}\label{supp_eq: FT_M^p}
    [F^T]_{(\bk, \beta), (\br, \beta)} M^p_{(\br, \beta), (\br', \beta')} F_{(\br', \beta'), (\bk', \beta')}\nonumber\\
    = \delta_{\bk, -\bk'} [M^p(\bk')]_{\beta,\beta'}
\end{align}
with 
\begin{equation}
    [M^p(\bk)]_{\beta,\beta'} = \sum_{\br}M^p_{(\br, \beta), (0, \beta')} e^{i\br\cdot \bk}.
\end{equation}
As such, we have
\begin{equation}
    b^\dagger_{(\br, \beta)} M^p_{(\br, \beta), (\br', \beta')} b^\dagger_{(\br', \beta')}  = b^\dagger_{(-\bk, \beta)} [M^p(\bk)]_{\beta,\beta'}b^\dagger_{(\bk, \beta')}.
\end{equation}
Notice that repeated indices are implicitly summed over. 

Below, we combine Eq.~\eqref{supp_eq: FT_M^p} with Eq.~\eqref{eq: real_space_relation} to obtain Eq.~\eqref{eq: k-space formula for M^p}. We examine
\begin{equation}
\begin{split}
    &F^T M^p F = F^T \mathscr{N}^{pp} F \\
    &+ F^T \mathscr{N}^{pv}[\mathscr{N}^{vv}+(M^\dagger)^{-1}]^{-1} (\mathscr{N}^{pv})^{T} F.
\end{split}
\end{equation}
The first term is very simple because $\mathscr{N}^{pp}$ is diagonal in $\br$ and translation-invariant (i.e., $\mathscr{N}^{pp}_{(\br,\beta),(\br', \beta')} = \delta_{\br,\br'}N^{pp}_{\beta,\beta'}$):
\begin{equation}
    [F^T \mathscr{N}^{pp} F ]_{(\bk, \beta),(\bk',\beta')}= \delta_{\bk, -\bk'} N^{pp}_{\beta,\beta'}.
\end{equation}
Similarly, as $\mathscr{N}^{pv}$ is diagonal in $\br$ and translation-invariant, for the second term we obtain
\begin{equation}
\begin{split}
    &F^T \mathscr{N}^{pv}[\mathscr{N}^{vv}+(M^\dagger)^{-1}]^{-1} (\mathscr{N}^{pv})^{T} F \\
    =&\mathscr{N}^{pv}[F^\dagger \mathscr{N}^{vv} F^* + ((F^\dagger M F^*)^\dagger)^{-1}]^{-1} (\mathscr{N}^{pv})^{T}
\end{split}
\end{equation}
where
\begin{equation}
    [F^\dagger \mathscr{N}^{vv} F^*]_{(\bk, \beta), (\bk',\beta')} = \delta_{\bk, -\bk'} N^{vv}_{\beta, \beta'}
\end{equation}
and
\begin{equation}
    [F^\dagger M F^*]_{(\bk, \beta), (\bk',\beta')} = \delta_{\bk, -\bk'}M(\bk)_{\beta, \beta'}.
\end{equation}
Using the real-space form of the $M$ matrix (describing the entangled pair states) given by Eq.~\eqref{eq: real-space M}, the explicit form of $M(\bk)$ is then given by Eq.~\eqref{eq: Mk_2D}. It is furthermore easy to check that $[(F^\dagger M F^*)^\dagger]^{-1}_{(\bk, \beta), (\bk',\beta')} = \delta_{\bk, -\bk'} [(M(\bk))^\dagger]^{-1}_{\beta, \beta'}=\delta_{\bk, -\bk'} M(\bk)_{\beta, \beta'}$. Combining all these pieces together, we arrive at
\begin{equation}
    M^p(\bk) = N^{pp}+N^{pv}[N^{vv}+M(\bk)]^{-1}(N^{pv})^T,
\end{equation}
where $N^{pp}$, $N^{pv}$ and $N^{vv}$ are the local site-independent TNS matrices. This gives Eq.~\eqref{eq: k-space formula for M^p} in the main text.

\section{Tuning $u(\bk)$ and $v(\bk)$ with 2D TNS}\label{appendix: Lagrange}
In this section, we describe the details of deriving the full forms of $u(\bk)$ and $v(\bk)$ as shown in ~\eqref{eq: det(N^vv+M) quadratic in B} and ~\eqref{eq: full v(k) to linear order in B} without $U(1)$ symmetry and ~\eqref{main text eq: full u(k) and v(k) for U(1)} with $U(1)$ symmetry as well as the proof by Lagrange interpolation that their coefficients can be independently tuned by the TNS parameters.
\subsection{Without $U(1)$}\label{subsec: Lagrange w/o U(1)}
Taking the $N^{vv}$ and $N^{pv}$ blocks to have the forms described in Sec.~\ref{subsec: 2D tuning u and v}, the calculation of $u(\bk)$ is done using Schur's formula for block matrices,
\begin{widetext}
\begin{equation}
\begin{split}
  u(\bk)=\det[N^{vv}+M(\bk)] &=\det\left[\begin{pNiceArray}[first-row,first-col]{cc|cc}
  & & & & \\
  &\mathbf{0}_{n_x} & N^{vv}_{LR}-e^{ik_x}\mathds{1}_{n_x} & N^{vv}_{LD} & N^{vv}_{LU} \\
  &N^{vv}_{RL}+e^{-ik_x}\mathds{1}_{n_x} & \mathbf{0}_{n_x} & N^{vv}_{RD} & N^{vv}_{RU} \\
  \hline
  &N^{vv}_{DL} & N^{vv}_{DR} & \mathbf{0}_{n_y} & N^{vv}_{DU}-e^{ik_y}\mathds{1}_{n_y} \\
  &N^{vv}_{UL} & N^{vv}_{UR} & N^{vv}_{UD}+e^{-ik_y}\mathds{1}_{n_y} & \mathbf{0}_{n_y}
\end{pNiceArray}\right]\\
    &=\det \begin{pNiceArray}[first-row,first-col]{c|c}
    & & \\
    & A & B \\
    \hline
    & -B^T & D
\end{pNiceArray} =\det(A)\det(D)\det(\mathds{1}_{2n_y}+D^{-1}B^T A^{-1} B),
\end{split}
\end{equation}
\end{widetext}
where $\{N^{vv}_{LD}, N^{vv}_{LU}, N^{vv}_{RD}, N^{vv}_{RU}\}$ make up the block B. Using the assumption that these matrices are small, the last term in the product is expanded to quadratic order in the $B$ block as
\begin{equation}
\begin{split}
    &\det(\mathds{1}_{2n_y}+D^{-1}B^T A^{-1} B) \\
    &\hspace{1cm}= 1+\Tr[D^{-1}B^T A^{-1} B] +\mathcal{O}(B^4).
\end{split}
\end{equation}
Carrying out the following matrix algebra results in ~\eqref{eq: det(N^vv+M) quadratic in B}.

For $v(\bk)=\det[N^{vv}+M(\bk)]M^p(\bk)$, the calculation is similar, using Schur's formula, but since $M^p(\bk)$ is already linear in $\{N^{vv}_{LD}, N^{vv}_{LU}, N^{vv}_{RD}, N^{vv}_{RU}\}$, we only need to keep the constant term from $\det[N^{vv}+M(\bk)]=\det(A)\det(D)+\mathcal{O}(B^2)$ to achieve the leading non-trivial order. For $M^p(\bk)=N^{pv}[N^{vv}+M(\bk)]^{-1}(N^{pv})^T$, we express the inverse term by again expanding for small $B$ this time at linear order
\begin{equation}
\begin{split}
    [N^{vv}&+M(\bk)]^{-1} = \begin{pNiceArray}{c|c}
  A & B \\
  \hline
  -B^T & D
\end{pNiceArray}^{-1} \\
&= \begin{pNiceArray}{c|c}
  A^{-1} & -A^{-1}BD^{-1} \\
  \hline
  D^{-1}B^T A^{-1} & D^{-1}
\end{pNiceArray} + \mathcal{O}(B^2)
\end{split}
\end{equation}
The resulting matrix algebra gives the full expression in ~\eqref{eq: full v(k) to linear order in B}.

We now show that the coefficients of $u(\bk)$ and $v(\bk)$ can be independently tuned by the TNS parameters, starting with $u(\bk)$, which is repeated below.
\begin{widetext}
    \begin{equation}\label{eq: u(k) appendix}
    \begin{split}
        &u(\bk) = \prod_{\lambda=1}^{n_x}\prod_{\mu=1}^{n_y}\; (e^{ik_x}-[N^{vv}_{LR}]_\lambda)(e^{-ik_x}-[N^{vv}_{LR}]_\lambda) (e^{ik_y}-[N^{vv}_{DU}]_\mu)(e^{-ik_y}-[N^{vv}_{DU}]_\mu) \\
        &\quad\quad\times \Bigg\{1+ \sum_{\lambda'=1}^{n_x}\sum_{\mu' = 1}^{n_y} \left[\frac{[N^{vv}_{LU}]_{\lambda'\mu'} [N^{vv}_{RD}]_{\lambda'\mu'}}{(e^{ik_x}-[N^{vv}_{LR}]_{\lambda'})(e^{ik_y}-[N^{vv}_{DU}]_{\mu'})}-\frac{[N^{vv}_{LD}]_{\lambda'\mu'} [N^{vv}_{RU}]_{\lambda'\mu'}}{(e^{ik_x}-[N^{vv}_{LR}]_{\lambda'})(e^{-ik_y}-[N^{vv}_{DU}]_{\mu'})}+ (\bk\rightarrow -\bk) \right]\Bigg\}.
    \end{split}
\end{equation}
\end{widetext}
We first note that since $u(\bk)$ and $v(\bk)$ can both be changed by an identical constant factor without affecting $M^p(\bk)=u(\bk)/v(\bk)$, the auxiliary parameter $\mathcal{U}$ which appears in $u(\bk) =\mathcal{U}\det[N^{vv}+M(\bk)]$ can be rescaled by the constant coefficient $u_\mathbf{0}$ to define a form with no constant term
\begin{equation}
\begin{split}
    \tilde{u}(\bk) &\equiv \frac{u(\bk)}{u_\mathbf{0}}-1=\frac{\mathcal{U}}{u_\mathbf{0}}\det[N^{vv}+M(\bk)] - 1 \\
    &= \sum_{\bmm\in \mathcal{L}(n_x,n_y)} \tilde{u}_\bmm\cos(\bmm\cdot \bk), \quad \tilde{u}_\bmm=\frac{u_{\bmm}}{u_\mathbf{0}}\in\mathbb{C}
\end{split}
\end{equation}

Thus, the parameter $\mathcal{U}$ is determined by $N^{vv}$ to guarantee $\tilde{u}(\bk)$ contains no constant term. Again, $\mathcal{L}(d_x,d_y)$ is the set of non-trivial lattice points $(\bmm\neq\mathbf{0})$ as defined in ~\eqref{eq: non-trivial lattice points} meaning there are $\abs{\mathcal{L}}=2n_xn_y+n_x+n_y$ coefficients $\{\tilde{u}_\bmm\}$ to be tuned by the TNS parameters. The appropriate TNS parameters needed to represent these coefficients are determined through Lagrange interpolation by the values of $\tilde{u}(\bk)$ at $\abs{\mathcal{L}}$ distinct points $\{\mathbf{k}_a\}$. These points are chosen based on the poles in ~\eqref{eq: u(k) appendix} as
\begin{widetext}
\begin{equation}\label{eq: choices of ki for Lagrange}
    \bk_a: \begin{cases}
        (e^{ik_{a,x}}\;,\; e^{ik_{a,y}})=([N^{vv}_{LR}]_a\;, \;y_0) \quad\quad&\text{for }\; a=1,2,..., n_x, \\
        (e^{ik_{a,x}}\;,\; e^{ik_{a,y}})=(x_0\;, \;[N^{vv}_{DU}]_{a-n_x}) \quad\quad&\text{for } \;a=n_x+1,n_x+2,..., n_x+n_y,\\
        (e^{ik_{a,x}}\;,\; e^{ik_{a,y}})=([N^{vv}_{LR}]_{\lambda_a}\;, \;[N^{vv}_{DU}]_{\mu_a}) \quad\quad&\text{for } \;a=n_x+n_y+1,..., n_x+n_y+n_x n_y,\\
        (e^{ik_{a,x}}\;,\; e^{-ik_{a,y}})=([N^{vv}_{LR}]_{\lambda'_a}\;, \;[N^{vv}_{DU}]_{\mu'_a}) \quad\quad&\text{for } \;a=n_x+n_y+n_xn_y+1,\;...\;, \abs{\mathcal{L}}.
    \end{cases}
\end{equation}
\end{widetext}
The values $x_0$ and $y_0$ are generic complex values chosen to differ from all parameters $[N^{vv}_{LR}]_{a=1,...,n_x}$ and $[N^{vv}_{DU}]_{a=n_x+1,...,n_x+n_y}$, and the subscripts $\{\lambda_a,\mu_a,\lambda'_a,\mu'_a\}$ are such that $(\lambda_a,\mu_a)$ and $(\lambda'_a,\mu'_a)$ run over all $n_xn_y$ combinations of $(1,1),(1,2),...,(n_x,n_y)$. This gives the total $\abs{\mathcal{L}}$ evaluations of $\tilde{u}(\bk_a)$ which are expressed in matrix form as
\begin{equation}
    \widetilde{u}(\bk_a) = \sum_{b=1}^{\abs{\mathcal{L}}} \mathcal{C}_{ab} \widetilde{u}_{\bmm_b},\quad \mathcal{C}_{ab} \equiv \cos(\bk_a\cdot\bmm_b).
\end{equation}
where $\mathcal{C}_{ab}$ is generally an invertible matrix due to our choices of $\{\mathbf{k}_i\}$. This allows us to determine the TP coefficients $\{\tilde{u}_{\bmm_b}\}$ from the evaluations $\{\tilde{u}(\bk_a)\}$ as $\tilde{u}_{\bmm_b}=\mathcal{C}^{-1}_{ba}\tilde{u}(\bk_a)$.

To independently tune each coefficient, we first define a reference point for the TNS values at $B=\mathbf{0}_{2n_x\times 2n_y}$ which results in $\tilde{u}(\bk_a)=-1$ for all $a$. Now, by varying individual TNS parameters in $N^{vv}$ while keeping $\{\bk_a\}$, and thus $\mathcal{C}_{ab}$, fixed, we can see how the coefficients $\tilde{u}_{\mathbf{m}_i}$ are changed as a result of the variation of $\tilde{u}(\bk)$. First, we see that varying $[N^{vv}_{LR}]_i+\delta [N^{vv}_{LR}]_i$ for any $i=1,...,n_x$ only varies the corresponding $i$-th component $\tilde{u}(\bk_i)$, and similarly, the following $n_y$ components are changed by the variations $\delta[N^{vv}_{DU}]_j$. Finally, as the values in $B$ are varied away from the reference point at $\mathbf{0}_{2n_x\times2n_y}$, the parameters $[N^{vv}_{LU}],\;[N^{vv}_{RD}],\; [N^{vv}_{LD}],\; \text{and} [N^{vv}_{RU}]$ take on non-zero values resulting in a subsequent $n_xn_y$ component variations each from the parameter combinations $[N^{vv}_{LU}]_{\lambda\mu}[N^{vv}_{RD}]_{\lambda\mu}$ and $[N^{vv}_{LD}]_{\lambda\mu}[N^{vv}_{RU}]_{\lambda\mu}$. Overall this results in the independent variations
\begin{widetext}
\begin{equation}
\begin{cases}
\tilde{u}(\bk_a)= -1+c_a \delta[N^{vv}_{LR}]_a \quad\quad &\text{for } a=1,2,...,n_x\\
\tilde{u}(\bk_a)= -1+c_a\delta [N^{vv}_{DU}]_{a-n_x} \quad\quad &\text{for } a=n_x+1,n_x+2,...,n_x+n_y\\
\tilde{u}(\bk_a)= -1+d_a[N^{vv}_{LU}]_{\lambda\mu}[N^{vv}_{RD}]_{\lambda\mu}\quad\quad &\text{for } a=n_x+n_y+1,...,n_x+n_y+n_xn_y\\
\tilde{u}(\bk_a)= -1+d_a[N^{vv}_{LD}]_{\lambda'\mu'}[N^{vv}_{RU}]_{\lambda'\mu'}\quad\quad &\text{for } a=n_x+n_y+n_xn_y+1,...,\abs{\mathcal{L}}\\
\end{cases}
\end{equation}
\end{widetext}
of each component of $\tilde{u}(\bk_a)$ using the $\abs{\mathcal{L}}=n_x+n_y+2n_xn_y$ TNS parameters in $N^{vv}$. Here, the coefficients $c_a$ and $d_a$ are each independent of the TNS parameter being varied at component $a$ so that the resulting component of $\tilde{u}(\bk_a)$ is linear in the associated variation for the $n_x+n_y$ components with coefficients $c_a$ and quadratic for the $2n_xn_y$ components with coefficients $d_a$. This allows all coefficients $\tilde{u}_{\mathbf{m}_b}=\mathcal{C}_{ba}^{-1}\tilde{u}(\bk_a)$ to be independently tuned. Along with the overall scaling parameter $\mathcal{U}$ used in defining the rescaled polynomial $\tilde{u}(\bk)$, this shows that all coefficients of $u(\bk)$ can be independently tuned by the TNS parameters, and upon using the Chevalley's theorem (see Sec.~\ref{subsec: Chevalley}), represent any TP.

For $v(\bk)$, the proof is nearly identical but without the need to define a rescaled polynomial as $v(\bk)$ can already be written in a similar form with no constant term. 
\begin{equation}
    v(\bk) = \sum_{\bmm \in \mathcal{L}(n_x,n_y)} v_\bmm \sin(\bmm\cdot \bk), \quad v_\bmm\in\mathbb{C}.
\end{equation}
The form of $v(\bk)$ is also repeated here for ease of reference. 
\begin{widetext}
\begin{equation}\label{eq: v(k) appendix}
\begin{split}
    v(\bk) = &\prod_{\lambda=1}^{n_x}\prod_{\mu=1}^{n_y}\;(e^{ik_x}-[N^{vv}_{LR}]_\lambda)(e^{-ik_x}-[N^{vv}_{LR}]_\lambda) (e^{ik_y}-[N^{vv}_{DU}]_\mu)(e^{-ik_y}-[N^{vv}_{DU}]_\mu) \\
    &\times\Big[ \sum_{\lambda'=1}^{n_x}\frac{-[N^{pv}_L]_{\lambda'}}{e^{ik_x}-[N^{vv}_{LR}]_{\lambda'}} + \sum_{\mu'=1}^{n_y}\frac{-[N^{pv}_D]_{\mu'}}{e^{ik_y}-[N^{vv}_{DU}]_{\mu'}} + \sum_{\lambda'=1}^{n_x}\sum_{\mu'=1}^{n_y}\frac{ [N^{pv}_L]_{\lambda'} [N^{vv}_{RD}]_{\lambda' \mu'}-[N^{pv}_{D}]_{\mu'}[N^{vv}_{LU}]_{\lambda'\mu'}}{(e^{ik_x}-[N^{vv}_{LR}]_{\lambda'})(e^{ik_y}-[N^{vv}_{DU}]_{\mu'})} \\
    &\quad\quad +\;\sum_{\lambda'=1}^{n_x}\sum_{\mu'=1}^{n_y}\frac{[N^{pv}_L]_{\lambda'}[N^{vv}_{RU}]_{\lambda'\mu'}[N^{pv}_D]_{\mu'}-[N^{vv}_{LD}]_{\lambda'\mu'}}{(e^{-ik_x}-[N^{vv}_{LR}]_{\lambda'})(e^{ik_y}-[N^{vv}_{DU}]_{\mu'})} - \;(\bk-\rightarrow - \bk)\;\Big].
\end{split}
\end{equation}
\end{widetext}
Since the same values appear in the poles of ~\eqref{eq: v(k) appendix}, we can use the same Lagrange points as in ~\eqref{eq: choices of ki for Lagrange}. The evaluation of the Lagrange points is expressed correspondingly allowing the coefficients of $v(\bk)$ to be solved for as
\begin{equation}\label{eq: relating v(k) to v_m}
     v_{\bmm_b}= \sum_{a=1}^{\abs{\mathcal{L}}} \mathcal{S}^{-1}_{ba} v(\bk_a), \quad \mathcal{S}_{ab}\equiv \sin(\bk_a\cdot\bmm_b).
\end{equation}
Again, the choices of $\bk_a$ guarantee that $\mathcal{S}_{ab}$ is invertible. This time, we define the reference point of the TNS values at $[N^{pv}_L]_a=1$ for all $a=1,...,n_x$ and $[N^{pv}_D]_a=1$ for all $a=1,...,n_y$ as well as $B=\mathbf{0}_{2n_x\times 2n_y}$. Thus at the reference point, $[N^{pv}]_a=1$ for all $a=1,...,2n_x+2n_y$ which results in $v(\bk)$ taking the form
\begin{widetext}
\begin{equation}
    \begin{split}
    v_r(\bk) = &\prod_{\lambda=1}^{n_x}\prod_{\mu=1}^{n_y}\;(e^{ik_x}-[N^{vv}_{LR}]_\lambda)(e^{-ik_x}-[N^{vv}_{LR}]_\lambda) (e^{ik_y}-[N^{vv}_{DU}]_\mu)(e^{-ik_y}-[N^{vv}_{DU}]_\mu) \\
    &\times\Big[ \sum_{\lambda'=1}^{n_x}\frac{-[N^{pv}_L]_{\lambda'}}{e^{ik_x}-[N^{vv}_{LR}]_{\lambda'}} + \sum_{\mu'=1}^{n_y}\frac{-[N^{pv}_D]_{\mu'}}{e^{ik_y}-[N^{vv}_{DU}]_{\mu'}}  - \;(\bk-\rightarrow - \bk)\;\Big].
\end{split}
\end{equation}
\end{widetext}
Then, in the same manner as before, we can independently vary components of $v(\bk_a)$ 
\begin{widetext}
\begin{equation}
\begin{cases}
v(\bk_a)=v_r(\bk_a)+e_a\delta [N^{pv}_L]_a \quad\quad &\text{for } a=1,2,...,n_x\\
v(\bk_a)=v_r(\bk_a)+ e_a\delta [N^{pv}_{D}]_{a-n_x} \quad\quad &\text{for } a=n_x+1,n_x+2,...,n_x+n_y\\
v(\bk_a)=v_r(\bk_a)+ f_a([N^{vv}_{RD}]_{\lambda\mu}-[N^{vv}_{LU}]_{\lambda\mu})\quad\quad &\text{for } a=n_x+n_y+1,...,n_x+n_y+n_xn_y\\
v(\bk_a)=v_r(\bk_a)+ f_a ([N^{vv}_{RU}]_{\lambda'\mu'}-[N^{vv}_{LD}]_{\lambda'\mu'})\quad\quad &\text{for } a=n_x+n_y+n_xn_y+1,...,\abs{\mathcal{L}}\\
\end{cases}
\end{equation}
\end{widetext}
Again, the coefficients $e_a$ and $f_a$ are independent of the variations at each component up to linear order in the variations. Thus, each coefficient of $v(\bk)$ is independently tuned via $v_{\mathbf{m}_b}=\mathcal{S}^{-1}_{ba}v(\bk_a)$, showing that the TNS parameters can be chosen to represent any TP (upon using the Chevalley's theorem, see Sec.~\ref{subsec: Chevalley}). Overall, this demonstrates that the $\mathcal{N}$ TNS parameters can be tuned so that both $u(\bk)$ and $v(\bk)$ represent any TP and thus any possible $M^p(\bk)=v(\bk)/u(\bk)$ can be represented by the TNS.

\subsection{With $U(1)$}\label{subsec: Lagrange w/ U(1)}

Similarly to the BCS case, and following Eq.~\eqref{eq: U(1) relation between the momentum-space wavefunction and the tensor network parameters}, we note that $u(\bk)$ is given by the determinant of $(N^{vv}+M(-\bk))$, up to a proportionality constant $\mathcal{U}$. Note that now, there is no longer any symmetry on the off-diagonal blocks for this matrix. We again break this matrix into blocks, $(N^{vv}+M(-\bk)) = \begin{pNiceArray}{c|c}
  A & B \\
  \hline
  C & D
\end{pNiceArray}$, and use Schur's formula to evaluate the determinant assuming the entries of $B$ and $C$ to be small:
\begin{equation}
\begin{split}
    \det \begin{pNiceArray}{c|c}
  A & B \\
  \hline
  C & D
\end{pNiceArray} = \det(A) \det(D) (1+&\Tr[D^{-1}CA^{-1}B])\\
&+\mathcal{O}(B^2C^2),
\end{split}
\end{equation}

Carrying out the evaluation of this expresssion gives the expression for $u(\bk)$,  up to a proportionality constant $\mathcal{U}$, as 
\begin{widetext}
\begin{equation}\label{eq: full u(k) for U(1)}
        \begin{split}
    u(\bk)&\propto\det (N^{vv}+M(-\bk))\\
    &= (-1)^{n_x+n_y}\prod_{\lambda=1}^{n_x}\prod_{\mu=1}^{n_y}\; (e^{-ik_x}-[N^{vv}_{LR}]_\lambda)(e^{+ik_x}+[N^{vv}_{RL}]_\lambda)(e^{-ik_y}-[N^{vv}_{DU}]_\mu)(e^{+ik_y}+[N^{vv}_{UD}]_\mu)\\
    &\hspace{1cm}\times \left(1+\sum_{\lambda=1}^{n_x}\sum_{\mu=1}^{n_y}\left[\frac{[N^{vv}_{RD}]_{\lambda\mu}[N^{vv}_{UL}]_{\mu\lambda}}{(e^{+ik_y}+[N^{vv}_{UD}]_\mu)(e^{+ik_x}+[N^{vv}_{RL}]_\lambda)}+\frac{(-1)^{n_x}[N^{vv}_{LD}]_{\lambda\mu}[N^{vv}_{UR}]_{\mu\lambda}}{(e^{+ik_y}+[N^{vv}_{UD}]_\mu)(e^{-ik_x}-[N^{vv}_{LR}]_\lambda)}\right.\right.\\
        &\hspace{3cm}\left.\left. +\frac{(-1)^{n_y}[N^{vv}_{RU}]_{\lambda\mu}[N^{vv}_{DL}]_{\mu\lambda}}{(e^{-ik_y}-[N^{vv}_{DU}]_\mu)(e^{+ik_x}+[N^{vv}_{RL}]_\lambda)}+\frac{(-1)^{n_x+n_y}[N^{vv}_{LU}]_{\lambda\mu}[N^{vv}_{DR}]_{\mu\lambda}}{(e^{-ik_y}-[N^{vv}_{DU}]_\mu)(e^{-ik_x}-[N^{vv}_{LR}]_\lambda)}\right]\right)
\end{split}
\end{equation}
\end{widetext}

Just as in the case without $U(1)$, we use the method of Lagrange interpolation to demonstrate that the following form can represent any TP of degree $(n_x,n_y)$:
\begin{equation}
    u(\bk)=\mathcal{U}\det(N^{vv}+M(-\bk))= \sum_{\mathbf{m}\in\mathcal{L}(n_x,n_y)}u_\mathbf{m}e^{i\mathbf{m}\cdot\bk}
\end{equation}
where $\mathcal{L}(n_x,n_y)$ here is the new set of lattice points defined in Eq.~\eqref{eq: U(1) lattice points}, with order $\abs{\mathcal{L}}=4n_xn_y+2n_x+2n_y+1$. To that end, we first define a rescaled form of $u(\bk)$ with the constant term removed
\begin{equation}
    \widetilde{u}(\bk) = \frac{u(\bk)}{u_\mathbf{0}}-1 = \sum_{\mathbf{m}\in\mathcal{L}(n_x,n_y)\setminus(0,0)}\widetilde{u}_\mathbf{m}e^{i\mathbf{m}\cdot\bk}
\end{equation}
Unlike the BCS case, the lattice points $\mathbf{m}$ and $-\mathbf{m}$ are no longer identified with each other resulting in twice as many coefficients as in the BCS case. Consequently, twice as many Lagrange points, $\{\bk_i\}$, must be specified which are chosen to be the poles of ~\eqref{eq: full u(k) for U(1)}. The change from $\bk$ to $-\bk$ also results in different Lagrange points than were used in the BCS case.
\begin{widetext}
\begin{equation}\label{eq: choices of ki for Lagrange U(1)}
    \bk_i: \begin{cases}
        (e^{-ik_{i,x}}\;,\; e^{-ik_{i,y}})=([N^{vv}_{LR}]_i\;, \;y_0) \quad\quad\text{for }\; i=1,2,..., n_x \\
        (e^{-ik_{i,x}}\;,\; e^{-ik_{i,y}})=(x_0\;, \;[N^{vv}_{DU}]_{i-n_x}) \quad\quad\text{for } \;i=n_x+1,n_x+2,..., n_x+n_y\\
        (e^{+ik_{i,x}}\;,\; e^{+ik_{i,y}})=([N^{vv}_{RL}]_{i-n_x-n_y}\;, \;y_0) \quad\quad\text{for }\; i=n_x+n_y+1,n_x+n_y+2,..., 2n_x+n_y \\
        (e^{+ik_{i,x}}\;,\; e^{+ik_{i,y}})=(x_0\;, \;[N^{vv}_{UD}]_{i-2n_x-n_y}) \quad\quad\text{for } \;i=2n_x+n_y+1,n_x+n_y+2,..., 2n_x+2n_y\\
        (e^{-ik_{i,x}}\;,\; e^{-ik_{i,y}})=([N^{vv}_{LR}]_{\lambda_i}\;, \;[N^{vv}_{DU}]_{\mu_i}) \quad\quad\text{for } \;i=2n_x+2n_y+1,..., 2n_x+2n_y+n_x n_y\\
        (e^{-ik_{i,x}}\;,\; e^{+ik_{i,y}})=([N^{vv}_{LR}]_{\lambda'_i}\;, \;[N^{vv}_{UD}]_{\mu'_i}) \quad\quad\text{for } \;i=2n_x+2n_y+1,..., 2n_x+2n_y+2n_x n_y\\
        (e^{+ik_{i,x}}\;,\; e^{-ik_{i,y}})=([N^{vv}_{RL}]_{\lambda''_i}\;, \;[N^{vv}_{DU}]_{\mu''_i}) \quad\quad\text{for } \;i=2n_x+2n_y+1,..., 2n_x+2n_y+3n_x n_y\\
        (e^{+ik_{i,x}}\;,\; e^{+ik_{i,y}})=([N^{vv}_{RL}]_{\lambda'''_i}\;, \;[N^{vv}_{UD}]_{\mu'''_i}) \quad\quad\text{for } \;i=2n_x+2n_y+3n_xn_y+1,\;...\;, \abs{\mathcal{L}}-1
    \end{cases}
\end{equation}
\end{widetext}
Here $x_0$ and $y_0$ are chosen to be generic complex values which differ from all of the TNS parameters $[N^{vv}_{LR}]_{i=1,2,...,n_x}$, $[N^{vv}_{DU}]_{i=1,2,...,n_y}$, etc. The subscripts $\{\lambda_i,\mu_i,\lambda'_i,\mu'_i,\lambda''_i,\mu''_i,\lambda'''_i,\mu'''_i\}$ indicate that as $i$ runs through the specified values, $(\lambda_i,\mu_i)$ runs through all $n_xn_y$ combinations: $(1,1), (1,2), ...,(n_x,n_y)$ and similarly for the other subscripts. The evaluation of $\widetilde{u}(\bk)$ at these Lagrange points is represented in terms of the vector of polynomial coefficients as
\begin{widetext}
\begin{equation}
    \begin{pNiceMatrix}[nullify-dots,xdots/line-style=loosely dotted]
    \widetilde{u}(\bk_1) \\ \widetilde{u}(\bk_2)\\ \widetilde{u}(\bk_3) \\ \Vdots \\ \Vdots\\ \widetilde{u}(\bk_{\abs{\mathcal{L}}-1})
\end{pNiceMatrix}=  \begin{pNiceMatrix}[nullify-dots,xdots/line-style=loosely dotted]
e^{i\bmm_1 \cdot \bk_1} & e^{i\bmm_2 \cdot \bk_1} & e^{i\bmm_3 \cdot \bk_1} & & \Cdots & e^{i\bmm_{\abs{\mathcal{L}}-1} \cdot \bk_1} \\
e^{i\bmm_1 \cdot \bk_2} & e^{i\bmm_2 \cdot \bk_2} & e^{i\bmm_3 \cdot \bk_2} & & \Cdots & e^{i\bmm_{\abs{\mathcal{L}}-1} \cdot \bk_2} \\
e^{i\bmm_1 \cdot \bk_3} & e^{i\bmm_2 \cdot \bk_3} & \Ddots &  & & \Vdots \\
 & \Vdots  &  &  & &  \\
\Vdots & & & & &  \\
e^{i\bmm_1 \cdot \bk_{\abs{\mathcal{L}}-1}} & e^{i\bmm_2 \cdot \bk_{\abs{\mathcal{L}}-1}} & \Cdots &  &  & e^{i\bmm_{\abs{\mathcal{L}}-1} \cdot \bk_{\abs{\mathcal{L}}-1}}
\end{pNiceMatrix} \begin{pNiceMatrix}[nullify-dots,xdots/line-style=loosely dotted]
    \widetilde{u}_{\bmm_1} \\ \widetilde{u}_{\bmm_2} \\ \widetilde{u}_{\bmm_3} \\ \Vdots \\ \Vdots\\ \widetilde{u}_{\bmm_{\abs{\mathcal{L}}-1}}
\end{pNiceMatrix} \equiv \mathcal{C}_{ij} \widetilde{u}_{\bmm_j}
\end{equation}
\end{widetext}
such that, with the above choice of the Lagrange points, the matrix $\mathcal{C}$ is generally invertible. We now choose as a reference point in the TNS parameter space the values $B=\mathbf{0}_{2n_x\times2n_y}$ and $C=\mathbf{0}_{2n_y\times2n_x}$. We can see from ~\eqref{eq: full u(k) for U(1)} that this results in the evaluation at the Lagrange points as
\begin{equation}
    \begin{pNiceMatrix}[nullify-dots,xdots/line-style=loosely dotted]
    \widetilde{u}_{\bmm_1} \\ \widetilde{u}_{\bmm_2} \\ \widetilde{u}_{\bmm_3} \\ \Vdots \\ \Vdots\\ \widetilde{u}_{\bmm_{\abs{\mathcal{L}}-1}}
\end{pNiceMatrix}= \mathcal{C}^{-1}\begin{pNiceMatrix}[nullify-dots,xdots/line-style=loosely dotted]
    \widetilde{u}(\bk_1) \\ \widetilde{u}(\bk_2)\\ \widetilde{u}(\bk_3) \\ \Vdots \\ \Vdots\\ \widetilde{u}(\bk_{\abs{\mathcal{L}}-1})
\end{pNiceMatrix}=   \mathcal{C}^{-1} \begin{pNiceMatrix}[nullify-dots,xdots/line-style=loosely dotted]
     -1\\ -1 \\ -1 \\ \Vdots \\ \Vdots\\ -1
\end{pNiceMatrix}.
\end{equation}
Now, by varying individual parameters of the TNS in $N^{vv}$ while keeping $\{\bk_i\}$, and thus $\mathcal{C}$ fixed, we can see how the coefficients $\widetilde{u}_{\mathbf{m}_i}$ are changed as a result of the variation of $\tilde{u}(\bk)$. First, varying either $[N^{vv}_{LR}]_\lambda$ or $[N^{vv}_{RL}]_\lambda \rightarrow [N^{vv}_{LR}]_\lambda+\delta_{x_i}$ or $[N^{vv}_{RL}]_\lambda+\delta_{x_i}$ for any $i=1...n_x$ only varies the $i$-th or $(n_x+n_y+i)$-th component of $\widetilde{u}(\bk)$
\begin{equation}
    \begin{pNiceMatrix}[nullify-dots,xdots/line-style=loosely dotted]
    \widetilde{u}(\bk_1) \\ \Vdots\\ \widetilde{u}(\bk_i) \\ \Vdots \\ \Vdots\\ \widetilde{u}(\bk_{\abs{\mathcal{L}}-1})
\end{pNiceMatrix} =  \begin{pNiceMatrix}[nullify-dots,xdots/line-style=loosely dotted]
    -1 \\ \Vdots\\ -1 \\ \Vdots \\ \Vdots\\ -1
\end{pNiceMatrix} + \begin{pNiceMatrix}[nullify-dots,xdots/line-style=loosely dotted]
    0 \\ \Vdots \\ \mathcal{O}(\delta x_i) \\ \Vdots \\ \Vdots\\ 0
\end{pNiceMatrix}
\end{equation}
and thus independently tunes $2n_x$ components. Similarly, varying $[N^{vv}_{UD}]_\mu$ or $[N^{vv}_{DU}]_\mu \rightarrow [N^{vv}_{UD}]_\mu+\delta_{y_j}$ or $[N^{vv}_{DU}]_\mu+\delta_{y_j}$ for any $j=1...n_y$ independently tunes the $(j+n_x)$ or $(j+2n_x+n_y)$-th parameters. And finally, each of the products $[N^{vv}_{RU}]_{\lambda\mu}[N^{vv}_{DL}]_{\mu\lambda}$, $[N^{vv}_{LU}]_{\lambda\mu}[N^{vv}_{DR}]_{\mu\lambda}$, $[N^{vv}_{RD}]_{\lambda\mu}[N^{vv}_{UL}]_{\mu\lambda}$, and $[N^{vv}_{LD}]_{\lambda\mu}[N^{vv}_{UR}]_{\mu\lambda}$ can be tuned using only the $n_xn_y$ parameters in each of the blocks of $B$ allowing tuning of the remaining $4n_xn_y$ components. All together, these $2n_x+2n_y+4n_xn_y$ TNS parameters in $N^{vv}$ make up the left side of Table~\ref{table: TNS parameters with U(1)} so that by independently tuning each component of $\widetilde{u}(\bk)$, this allows all coefficients $\widetilde{u}_{\mathbf{m}_i}=\mathcal{C}_{ij}^{-1}\widetilde{u}(\bk_j)$ to be independently tuned. Along with the overall scaling parameter $\mathcal{U}$ used in defining the rescaled polynomial $\widetilde{u}(\bk)$, this shows that all coefficients of $u(\bk)$ can be independently tuned and thus the TNS can represent any TP (upon using the Chevalley's theorem, see Sec.~\ref{subsec: Chevalley}).

Since $v(\bk)$ is a TP of the same form as $u(\bk)$, much of the argument of tuning its coefficients is identical to that of $u(\bk)$. Recalling that $v(\bk)=u(\bk)M^p(\bk)$, we evaluate $M^p(\bk)$ for the same perturbative expansion for small $N^{vv}_{xy}$ and $N^{vv}_{yx}$.
\begin{equation}
\begin{split}
    (N^{vv}+&M(-\bk))^{-1} = \begin{pNiceArray}{c|c}
  A & B \\
  \hline
  C & D
\end{pNiceArray}^{-1}\\
&=\begin{pNiceArray}{c|c}
  A^{-1} & -A^{-1}BD^{-1} \\
  \hline
  -D^{-1}CA^{-1} & D^{-1}
\end{pNiceArray}+\mathcal{O}(BC)\\
\end{split}
\end{equation}
We keep only the lowest order term in $B$ and $C$, and we introduce condensed notation to represent the relevant combinations of parameters. Specifically, similarly to ~\eqref{eq: restricted form of N^{pv}}, we assert the form of $N^{pv}$ to have $N^{pv}_-=(\ [1]_{n_x}\ |\ [1]_{n_x}\ |\ [1]_{n_y}\ |\ [1]_{n_y}\ )$ and define the following parameters 
\begin{widetext}
\begin{alignat*}{2}
        &a_{\lambda\in[n_x]} = [N^{pv}_{+L}]_\lambda\ ; \hspace{2cm} &&\tilde{a}_{\lambda\in[n_x]}=[N^{pv}_{+R}]_\lambda \ ;\\
        &b_{\mu\in[n_y]} = [N^{pv}_{+D}]_\mu\ ;  \hspace{2cm} &&\tilde{b}_{\mu\in[n_y]}=[N^{pv}_{+U}]_\mu \\
        &c_{\lambda\mu} = b_\mu[N^{vv}_{UR}]_{\mu\lambda} + \tilde{a}_\lambda[N^{vv}_{LD}]_{\lambda\mu}\ ;  \hspace{0.5cm} &&\tilde{c}_{\lambda\mu} = \tilde{b}_\mu[N^{vv}_{DR}]_{\mu\lambda} + \tilde{a}_\lambda[N^{vv}_{LU}]_{\lambda\mu}\ ;\\
        &d_{\lambda\mu} = b_\mu[N^{vv}_{UL}]_{\mu\lambda} + a_\lambda[N^{vv}_{RD}]_{\lambda\mu}\ ;\hspace{0.5cm} &&\tilde{d}_{\lambda\mu}=\tilde{b}_\mu[N^{vv}_{DL}]_{\mu\lambda} + a_\lambda[N^{vv}_{RU}]_{\lambda\mu}
\end{alignat*}
\end{widetext}
These allow the final form of $v(\bk)$ which appears in Eq.~\eqref{main text eq: full u(k) and v(k) for U(1)} to be rewritten in terms of the simplified parameters
\begin{widetext}
    \begin{equation}\label{v(k) U(1}
    \begin{split}
        v&(\bk) = (-1)^{n_x+n_y}\prod_{\lambda=1}^{n_x}\prod_{\mu=1}^{n_y}\; (e^{-ik_x}-[N^{vv}_{LR}]_\lambda)(e^{+ik_x}+[N^{vv}_{RL}]_\lambda)(e^{-ik_y}-[N^{vv}_{DU}]_\mu)(e^{+ik_y}+[N^{vv}_{UD}]_\mu)\\
        \times&\left\{N^{pp}+\sum_{\lambda=1}^{n_x}\left[\frac{\tilde{a}_\lambda}{([N^{vv}_{LR}]_\lambda-e^{-ik_x})}+\frac{a_\lambda}{([N^{vv}_{RL}]_\lambda+e^{+ik_x})}\right] + \sum_{\mu=1}^{n_y}\left[\frac{\tilde{b}_\mu}{([N^{vv}_{DU}]_\mu-e^{-ik_y})} + \frac{b_\mu}{([N^{vv}_{UD}]_\mu+e^{+ik_y})}\right] \right.\\
        &\left.- \sum_{\lambda=1}^{n_x}\sum_{\mu=1}^{n_y}\left[ \frac{c_{\lambda\mu}}{([N^{vv}_{UD}]_\mu + e^{+ik_y})([N^{vv}_{LR}]_\lambda - e^{-ik_x})} + \frac{\tilde{c}_{\lambda\mu}}{([N^{vv}_{DU}]_\mu - e^{-ik_y})([N^{vv}_{LR}]_\lambda - e^{-ik_x})} \right.\right. \\
        &\hspace{1.25cm}\left.\left. + \frac{d_{\lambda\mu}}{([N^{vv}_{UD}]_\mu + e^{+ik_y})([N^{vv}_{RL}]_\lambda + e^{+ik_x})}+\frac{\tilde{d}_{\lambda\mu}}{([N^{vv}_{DU}]_\mu - e^{-ik_y})([N^{vv}_{RL}]_\lambda + e^{+ik_x})} \right]\right\} \\
    \end{split}
\end{equation}
\end{widetext}
Again defining the rescaled form of $v(\bk)$ as
\begin{equation}
    \widetilde{v}(\bk) = \frac{v(\bk)}{v_\mathbf{0}}-1=\sum_{\mathbf{m}\in\mathcal{L}(n_x,n_y)\setminus(0,0)}\widetilde{v}_\mathbf{m} e^{i\mathbf{m}\cdot\bk}
\end{equation}
we use the same set of Lagrange points as before and thus the same $\mathcal{C}$ matrix to write the coefficients as $\widetilde{v}_{\mathbf{m_i}}=\mathcal{C}^{-1}_{ij}\widetilde{v}(\bk_j)$. In this case, the reference point in parameter space is where all of the simplified parameters $\{a_\lambda,\tilde{a}_\lambda, b_\mu, \tilde{b}_\mu, c_{\lambda\mu}, \tilde{c}_{\lambda\mu}, d_{\lambda\mu}, \tilde{d}_{\lambda\mu}\}=0$. By the same procedure as before, this allows $2n_x$ coefficients to be tuned by varying each of $a_\lambda$ or $\tilde{a}_\lambda\rightarrow a_\lambda+\delta_{x_i}$ or $\tilde{a}_\lambda+\delta_{x_i}$ and a further $2n_y$ coefficients by varying $b_\mu$ or $\tilde{b}_\mu\rightarrow b_\mu+\delta_{y_i}$ or $\tilde{b}_\mu+\delta_{y_i}$. The remaining $4n_xn_y$ parameters are tuned by the variation of the parameters $c_{\lambda\mu}, \tilde{c}_{\lambda\mu}, d_{\lambda\mu}$, and $\tilde{d}_{\lambda\mu}$. Recall that the parameters in the $N^{vv}_{LD}, N^{vv}_{LU}, N^{vv}_{RD}$, and $N^{vv}_{RU}$ matrix blocks were already fixed in tuning the coefficients of $u(\bk)$, but these $4n_xn_y$ parameters can still be varied using the parameters in the $N^{vv}_{UR}, N^{vv}_{DR}, N^{vv}_{UL}$, and $N^{vv}_{DL}$ matrix blocks. These simplified parameters make up the right side of Table~\ref{table: TNS parameters with U(1)} again showing that all coefficients of $v(\bk)$ can be independently tuned, and by Chevalley's theorem (see Sec.~\ref{subsec: Chevalley}), this allows the TNS to represent any TP for $v(\bk)$.  

\section{Counting argument for the bond dimension bound with $n_p>2$ bands}\label{sec: appendix, counting for np > 2}
Let us now explore the more general case of having $n_p$ bands (orbitals) but filling only 1 band. The Gaussian state is represented generally as
\begin{equation}\label{eq: U(1)-symmetric Gaussian state, generic n_p}
    \ket{\psi_p} \propto \prod_\bk (1+\sum_{\sigma=2}^{n_p}M^p_\sigma(\bk)c^\dagger_{(\bk,\sigma)}c_{(\bk,1)})\ket{10...0}_p,
\end{equation}
where the reference state $\ket{10...0}_p$ corresponds to filling the first orbital and leaving all other orbitals $(\sigma=2,...,n_p)$ empty for all $\bk$.  To match with the TNS structure, we again assume $M^p_\sigma(\bk)$ is a ratio of TPs: 
\begin{equation}
\begin{split}
    &M^p_\sigma(\bk) = \frac{v_\sigma(\bk)}{u(\bk)} \\
    &= \frac{\sum_{m_x=-\mathcal{M}_x}^{\mathcal{M}_x}\sum_{m_y=-\mathcal{M}_y}^{\mathcal{M}_y} v^{(m_x,m_y)}_\sigma e^{im_x k_x+im_y k_y}}{\sum_{m_x=-\mathcal{M}_x}^{\mathcal{M}_x}\sum_{m_y=-\mathcal{M}_y}^{\mathcal{M}_y} u^{(m_x,m_y)}e^{im_x k_x+im_y k_y}}.
\end{split}
\end{equation}
Counting the number of independent TP coefficients, we see that the physical state $\ket{\psi_p}$ is specified by 
\begin{equation}
    \mathcal{N}_p =4\mathcal{M}_x\mathcal{M}_yn_p+2(\mathcal{M}_x+\mathcal{M}_y)n_p+n_p-1
\end{equation}
number of independent complex parameters (with $-1$ above coming from the over-counting due to overall rescaling of $u$ and $v_\sigma$'s). Notice that $\ket{\psi_p} \propto\prod_\bk d^\dagger_\bk \ket{0}_p \propto \prod_\br d^\dagger_\br \ket{0}_p$, where $d^\dagger_\bk = u(\bk)c^\dagger_{(\bk,1)}+\sum_{\sigma=2}^{n_p}v_\sigma(\bk)c^\dagger_{(\bk,\sigma)}$, and $d^\dagger_\br$ is the Fourier transform of $d^\dagger_\bk$ which creates a compactly supported Wannier-type orbital of size $(2\mathcal{M}_{x}+1)\times (2\mathcal{M}_{y}+1)$. From the Schimdt decomposition of $\ket{\psi_p} \propto \prod_\br d^\dagger_\br \ket{0}_p$, it is straightforward to see that the Schmidt rank (or R\'enyi-0 entanglemet entropy) for a vertical (horizontal) partition cut of length $l_y$ ($l_x$) is upper bounded by $2\mathcal{M}_{x}l_y$ ($2\mathcal{M}_{y}l_x$).

On the other hand, following Sec.~\ref{sec: U(1) TNS review}, consider a TNS with $n_{x/y}$ number of virtual complex fermions in the $(\pm)$-sector on each horizontal/vertical bond, and 1 physical fermion per site in the $(+)$-sector and $n_p-1$ physical fermions per site in the $(-)$-sector. To preserve $U(1)$-symmetry, the local tensor contains couplings only between $(+)$ and $(-)$ sectors, and hence there are $(2n_x+2n_y+1)\times (2n_x+2n_y+n_p-1)$ number of TNS parameters. Subtracting off $4n_x^2+4n_y^2$ number of gauge redundancies, we conclude that
\begin{equation}
    \mathcal{N}_{\rm TNS} = 8n_xn_y+2(n_x+n_y)n_p+n_p-1
\end{equation}
is the number of independent TNS parameters. To have sufficient TNS expressivity, we require $\mathcal{N}_{\rm TNS}\geq \mathcal{N}_p$, or equivalently
\begin{equation}
     4n_xn_y+(n_x+n_y)n_p \geq 2\mathcal{M}_x\mathcal{M}_yn_p+(\mathcal{M}_x+\mathcal{M}_y)n_p
\end{equation}
which can be satisfied by picking
\begin{equation}\label{eq: app, bond dimension bound}
    n_x \geq \mathcal{M}_x\sqrt{\frac{n_p}{2}} \quad \text{and}\quad n_y\geq\mathcal{M}_y\sqrt{\frac{n_p}{2}}.
\end{equation}
While here we have focused on cases with one occupied band in the $n_p$-band system, it is natural to expect that $\mathcal{N}_p$ (the number of parameters to specify $\ket{\psi_p}$) would generally be higher if more bands are occupied, as there is a more complex system to characterize. The bound given in Eq.~\eqref{eq: app, bond dimension bound} is thus a conservative estimation, while the actual required $n_{x,y}$ may be even larger. This demonstrates that for $n_p>2$, the minimal bond dimension (given by $2^{2n_{x,y}}$ for the $U(1)$-symmetric GfTNS) is not solely determined by the entanglement area-law bound given by $\mathcal{M}_{x,y}$.

\bibliography{ref.bib}
\end{document}